\documentclass[11pt]{article}
 \usepackage{epsfig} \parskip 5pt plus
1pt \textheight 22cm \textwidth 15.5cm \oddsidemargin 0.0cm
\newcommand{\nn}{\nonumber}

\newcommand{\be}{\begin{equation}} \newcommand{\ee}{\end{equation}}
\newcommand{\bea}{\begin{eqnarray}} \newcommand{\eea}{\end{eqnarray}}

\def\beq{\begin{equation}}
\def\eeq{\end{equation}}

\newcommand{\tetaot}{\mbox{$\theta_{13}$}}
\newcommand{\tetaotp}{\mbox{$\theta^{'}_{13}$}}
\newcommand{\deltap}{\mbox{$\delta^{'}$}}
\newcommand{\tetatt}{\mbox{$\theta_{23}$}}

\newcommand{\sol}{\mbox{${\Delta m^2_{12} L \over 4 E}$}}
\newcommand{\atmos}{\mbox{${\Delta m^2_{13} L \over 4 E}$}}
\newcommand{\atmosdos}{\mbox{${\Delta m^2_{13} L \over 2 E}$}}

\newcommand{\enu}{\mbox{$\langle E_\nu \rangle$}}
\begin{document}

%%%%%%%%%%%%%%%%%%%%%%%%%%%%%%%%%%%%%%%%%%%%%%%%%%%%%%%%%%%%%%%%%%%%%%%%%%%%%%%%
%
\thispagestyle{empty}
\begin{flushright}
%{\tt hep-ph/}\\
{CERN-TH/2002-148}\\
{FTUAM-02-18}\\
{IFT-UAM/CSIC-02-26} \\
{FTUV-02-0704}\\
{IFIC/02-27}
\end{flushright}
\vspace*{1cm}
\begin{center}
{\Large{\bf Superbeams plus Neutrino Factory: the golden path to leptonic CP violation} }\\
\vspace{.5cm}
J. Burguet-Castell$^{\rm a,}$, M.B. Gavela$^{\rm b,}$\footnote{gavela@delta.ft.uam.es}, J.J. G\'omez-Cadenas$^{\rm a,c}$\footnote{gomez@hal.ific.uv.es},
P. Hern\'andez$^{\rm c,}$\footnote{pilar.hernandez@cern.ch. On leave 
from Dept. de F\'{\i}sica Te\'orica, Universidad de Valencia.}, 
O. Mena$^{\rm b,}$\footnote{mena@delta.ft.uam.es}
 
\vspace*{1cm}
$^{\rm a}$ Dept. de F\'{\i}sica At\'omica y Nuclear and IFIC, Universidad de Valencia, Spain \\
$^{\rm b}$ Dept. de F\'{\i}sica Te\'orica, Univ. Aut\'onoma de
Madrid, Spain \\
$^{\rm c}$ CERN, 1211 Geneva 23, Switzerland

\vspace{.3cm}

%

%%%%%%%%%%%%%%%%%%%%%%%%%%%%%%%%%%%%%%%%%%%%%%%%%%%%%%%%%%%%%%%%%%%%%%%%%%%

\begin{abstract}
\noindent

 Superbeams (SB) and Neutrino Factories (NF) are not alternative facilities for exploring neutrino oscillation 
physics, but successive steps. The correct strategy is to contemplate the combination of their expected 
physics results. We show its important potential on the disappearance of fake degenerate 
solutions in the simultaneous measurement of $\theta_{13}$ and leptonic CP violation.
Intrinsic, sign($\Delta m_{13}^2$) and $\theta_{23}$ degeneracies are 
shown to be extensively eliminated when the results from one NF baseline and a SB 
facility are combined. A key point is the different average neutrino energy and 
baseline of the facilities. 
For values of $\theta_{13}$ near its present limit, the short NF baseline, e.g. $L=732$ km, 
becomes, after such a combination, a very interesting distance.
For smaller  $\theta_{13}$, an intermediate NF baseline of $O(3000$km) is still required.

\end{abstract}

\end{center}
%

%%%%%%%%%%%%%%%%%%%%%%%%%%%%%%%%%%%%%%%%%%%%%%%%%%%%%%%%%%%%%%%%%%%%%%%%%%%%%%%%
%

\pagestyle{plain} 
\setcounter{page}{1}

\section{Introduction}

  Recent data \cite{sno} strongly favour the large mixing angle solution (LMA-MSW) \cite{MSW} to the 
solar neutrino deficit \cite{solar}.
This, if confirmed, is very good news as regards the prospects of discovering leptonic 
CP violation. The measurement of the angle 
$\theta_{13}$ and the CP-odd
phase $\delta$ would be possible at a superbeam (SB)
facility \cite{firstsb,mezzetosb,sb}  or/and a neutrino factory (NF) \cite{reviewnf}, mainly 
through $\nu_\mu \leftrightarrow \nu_e$ and  $\bar \nu_\mu \leftrightarrow \bar \nu_e$ oscillations, provided  
$\theta_{13}$ is not too small. How ``small'' is one of the points to be quantified below.

 The development of a neutrino factory requires, by design, the essentials of a superbeam facility  
as an intermediate step.
Although the ultimate precision and discovery goals in neutrino oscillation 
physics may only be attained with a neutrino factory from muon storage rings, those ``for free'' superbeam results can 
already lead to significant progress in central physics issues such as those 
mentioned above.

Superbeams and neutrino factory are thus not alternative options, but successive steps.
In this perspective, the correct analysis strategy is to contemplate the combination of their expected
physics results. We will take such a step here and show its important impact, in particular on the 
disappearance of fake solutions in the simultaneous measurement of $\tetaot$ and $\delta$.

It was shown in ref.~\cite{burguet} that there exists generically, at a given
(anti)neutrino energy and fixed baseline, 
a second value of the set ($\tetaotp,\deltap$) that gives the same 
oscillation probabilities for neutrinos and antineutrinos as
the true value chosen by nature. In what follows, these fake solutions 
will be dubbed  {\it intrinsic degeneracies}.

More recently it has been pointed out \cite{bargerdeg} that other fake solutions might appear from 
unresolved degeneracies in two other oscillation parameters: 
\begin{itemize}
\item the sign of $\Delta m_{13}^2$, 

\item $\theta_{23}$, upon the exchange $\theta_{23} \leftrightarrow \pi/2 - \theta_{23}$ for $\theta_{23} \ne \pi/4$. 

\end{itemize}

It is not expected that these degeneracies will be resolved
before the time of the SB/NF operation. However, the subleading transitions
$\nu_e \leftrightarrow \nu_\mu$, from which the parameters $\tetaot$ and $\delta$
can be measured, are sensitive to these discrete
ambiguities. A complete analysis of the sensitivity to the set 
$(\tetaot, \delta)$ should therefore assume
that  sign$(\Delta m^2_{13})$ can be either positive or negative and 
$\theta_{23} >$ or $< \pi/4$. If a wrong choice of these possibilities 
cannot fit the data, the ambiguities will be resolved, else they will generically give 
rise to new fake solutions for the parameters $\tetaot$ and $\delta$.

Strategies to eliminate some of the fake solutions have previously been discussed, advocating the combination of 
different baselines \cite{burguet}, an improved experimental technique allowing the measurement of the neutrino 
energy with good precision\cite{lindner}, the supplementary detection of $\nu_e \rightarrow \nu_\tau$ channels \cite{andrea} 
and a cluster of detectors at
a superbeam facility located at different off-axis angles, so as to have 
different $\langle E\rangle$ \cite{bargernew}.

We consider the three types of degeneracies as they would appear in the analysis from the data taken 
at  a neutrino factory and at an
associated superbeam. We then show the potential  of combining their results. For the NF 
we consider the experimental setup presented in \cite{lmd,golden,burguet}. The parent $\mu^\pm$ energy 
is $50$ GeV and the reference baselines considered will be 732 and 2810 km. As the  
SB facility we consider the design proposed for the CERN SPL \cite{sb2}, which could be the 
initial step of a NF based at CERN. The average energy is $\langle E_\nu \rangle = 0.25$ GeV and the baseline is
 $L=130$ km (CERN-Fr\'ejus). The fluxes and detector systematics have been discussed in 
\cite{sb}. In the detailed computations of this paper we will not consider other types of 
superbeams, such as JHF \cite{jhf}, nor the case of beta-beams \cite{zuchelli}, although we
will also discuss the potential of  
these alternative experimental setups in this context. 

Whenever it is not specified otherwise, we take the following central values
for the oscillation parameters: $\sin 2 \theta_{12} \;. \Delta m^2_{12} = 10^{-4}$eV$^2$ with $\Delta m^2_{12}>0$, 
$|\Delta m^2_{13}| = 3\times 10^{-3}$ eV$^2$ and $\sin 2 \theta_{23} = 1$. Some implications of making them vary      
within their currently allowed ranges will be discussed as well. The errors expected in the knowledge of these parameters, as well as the error in the matter profile of the Earth will not be included.  We previously studied 
their impact in ref.~\cite{burguet} and they are not expected to change the conclusions significantly. 

\section{ Degenerate solutions}

In this study we will consider in detail the measurement of the subleading 
probabilities $P_{\nu_ e\nu_\mu ( \bar \nu_e \bar \nu_\mu )}$ in the NF complex, through the detection of ``wrong-sign'' muons, 
and $P_{\nu_\mu\nu_e ( \bar \nu_\mu \bar \nu_e )}$ in the SB facility, through the detection of electrons/positrons. 

For the energies and baselines under discussion, the oscillation probabilities in vacuum are accurately given 
by (for more details see \cite{golden}): 
\bea
P_{\nu_ e\nu_\mu ( \bar \nu_e \bar \nu_\mu ) } & = & 
s_{23}^2 \, \sin^2 2 \tetaot \, \sin^2 \left(\atmos\right)  + 
c_{23}^2 \, \sin^2 2 \theta_{12} \, \left( \sol\right)^2 \nn \\
& + & \tilde J \, \cos \left ( \delta \mp \atmos \right ) \;
 \sol \sin \atmos \,\,,  
\label{vacexpand} 
\eea
where $\tilde J$ is defined as 
\be
\tilde J \equiv \cos \theta_{13} \; \sin 2 \theta_{13}\; \sin 2 \theta_{23}\;
\sin 2 \theta_{12}.
\ee
The vacuum approximation should be excellent as regards the SB scenario with a baseline of hundreds of kilometers, 
while in practice it also gives a good indication for the results at the short ($732$ km) and intermediate ($2000$--$3000$km) baselines 
of a NF.
Of course all the fits  below include the exact formulae for the 
probabilities, including matter effects.

As in ref. \cite{burguet}, we will denote the three terms 
in eq.~(\ref{vacexpand}), atmospheric, solar and interference, by $P^{atm}_{\nu ( \bar \nu ) }$, $P^{sol}$ and $P^{inter}_{\nu ( \bar \nu) }$, respectively. When $\theta_{13}$ is 
relatively large or $|\Delta m^2_{12}|$ small, the probability is dominated
by the atmospheric term. We will 
refer to this situation as the atmospheric regime. Conversely, when 
$\theta_{13}$ is very small or $|\Delta m^2_{12}|$ large (with respect to 
$\enu/L$ and $|\Delta m^2_{13}|$), the solar 
term dominates $P^{sol} \gg P^{atm}_{\nu (\bar \nu )}$. This is the solar
regime. The interference term, which contains the information on the CP-violating phase $\delta$, can never dominate since 
\bea
 |P^{inter}_{\nu(\bar \nu)}| \leq P^{atm}_{\nu ( \bar \nu ) } + P^{sol}.
\eea
It is precisely in the transition between the two regimes where the 
interference term becomes larger in relative terms. There the corresponding 
CP-odd asymmetry becomes maximal and independent of the precise value of the two small parameters: 
$\Delta m^2_{12}$ and $\theta_{13}$ \cite{burguet}. 
As an indication, for the solar parameters $\sin 2\theta_{12} \cdot \Delta m^2_{12} = 10^{-4}$ eV$^2$, the transition 
is at $ \tetaot \simeq 1^\circ$ for the 
NF setups considered here and $\tetaot \simeq 2^\circ$ for the SPL-SB facility.

\section{Intrinsic degeneracies}

In a previous work \cite{burguet}  we uncovered, at fixed neutrino
energy and baseline, the existence of degenerate solutions in
the ($\theta_{13}$, $\delta$) plane for fixed values of the oscillation 
probabilities $\nu_e(\bar{\nu}_e)\rightarrow\nu_\mu(\bar{\nu}_\mu)$. 
If ($\theta_{13}$, $\delta$) are the values chosen by nature, the conditions
\vspace{-0.75cm}
\begin{center}
\begin{equation}
\left.\matrix{
P_{\nu_e \nu_\mu} (\theta^{'}_{13}, \delta^{'}) = P_{\nu_e \nu_\mu}
(\theta_{13}, \delta)\nonumber \cr 
P_{\bar \nu_e \bar \nu_\mu} (\theta^{'}_{13}, \delta^{'}) = P_{\bar \nu_e
\bar \nu_\mu} (\theta_{13}, \delta)}
\right \}
\label{equalburguet}
\end{equation}
\end{center}
can be generically satisfied by another set ($\theta^{'}_{13}$, $\delta^{'}$). 
Using the approximate formulae of eq.~(\ref{vacexpand}), it is easy to 
find the expression for these {\it intrinsic} degeneracies deep in the atmospheric and solar regimes. In ref. \cite{burguet}
the general results including matter effects  were presented. Here we just 
recall the solutions in vacuum, which are simpler and accurately describe
the new situation considered here, that of the superbeams.

For $\theta_{13}$ sufficiently large and in the vacuum approximation, apart from  the true solution,  
$\deltap=\delta$ and $\tetaotp=\tetaot$, there is a fake one at 
\bea
\delta^{'}&\simeq&\pi -\delta,\nonumber\\
\theta^{'}_{13}&\simeq & \theta_{13} +
\cos\delta\ \sin 2\theta_{12} \;\sol      \cot \theta_{23} \cot\left(\atmos\right) \; \;.
\label{atmburguet}
\eea 
Note that for the values $\delta =-90^\circ, 90^\circ$, the two solutions degenerate into one. 
Typically  $\cot\left(\atmos\right)$ has on average opposite signs for the proposed SB and the NF setups\footnote{Clearly the parameters for these setups are not fixed yet and might be modified conveniently in the final designs.}, for $\Delta m^2_{13}=0.003$ eV$^2$:
\begin{center}
\[\begin{array}{c|c|c|c}
&\langle E\rangle ({\rm GeV}) & L ({\rm km}) & \cot\left({\Delta m^2_{13} L \over 4 E}\right)\\
\hline
{\rm SB-SPL}&0.25&130&{-}0.43\\
\rm{JHF-off-axis}&0.7&295&{-}0.03\\
\rm{{NF}}@732&30&732&{+}10.7\\
\rm{{NF}}@2810&30&2810&{+}2.68\\
\beta{\rm-beam}&0.35&130&{+}0.17\\
\hline
\end{array}\]
\end{center}

When $\theta_{13} \rightarrow 0$ and in the vacuum approximation, the intrinsic degeneracy is 
independent\footnote{All throughout this paper we will only consider those fake solutions which 
lie inside the experimentally allowed range for $\tetaot$.}
 of $\delta$:

\vspace{-1.0cm}
 \begin{center}
\begin{equation}
\left.\matrix{
\textrm{if}\ \cot\left(\atmos\right)  > 0\  \textrm{then} \ \delta^{'}\simeq\ \pi \cr
\textrm{if}\ \cot\left(\atmos\right)  < 0\  \textrm{then} \  \delta^{'}\simeq\ 0}
\right \} \;\;\; \theta^{'}_{13}\simeq \sin 2 \theta_{12}\  \sol \ |\cot\theta_{23}\ \cot\left(\atmos\right) \ |.
 \label{solarburguet}
\end{equation}
\end{center}

This solution was named $\tetaot =0$-mimicking solution and occurs because there is a value of $\tetaotp$ for which there 
is an exact cancellation of the atmospheric and
interference terms in both the neutrino and antineutrino probabilities 
simultaneously, with $\sin \delta'=0$. 
\begin{figure}[t]
\begin{center}
\epsfig{file=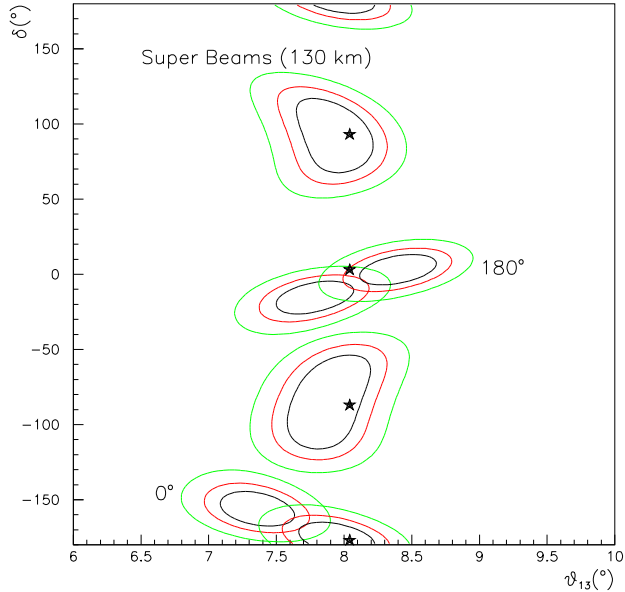, width=9cm} 
\end{center}
\caption{\it{ Fits to given true (nature) solutions and their intrinsic degenerate solutions at a SB facility. 
The $68.5\%$, $90\%$ and $99\%$ contours are depicted, for
four central values of $\delta=- 90^\circ, 0^\circ, 90^\circ,
180^\circ$ and for $\theta_{13}= 8^\circ$.}}
\label{sbeams}
\end{figure}

Figure~\ref{sbeams} shows the results of measuring $(\tetaot, \delta)$ at the SPL-SB 
facility, for $\tetaot = 8^\circ $ and the central values 
of $\delta=-180, -90, 90, 180^\circ$.  
The intrinsic degeneracies clearly appear
and are well described by eqs.~(\ref{atmburguet}). The details of the analysis can 
be found in ref. \cite{sb}. We simply stress here that the analysis is based on the total number of electron/positron events, so 
we do not assume that the 
neutrino energy can be reconstructed. Our results about the NF can be found in ref. \cite{burguet}, where it was 
shown that these intrinsic degeneracies also could not be resolved in a single baseline, in spite of the spectral 
information, although they did get eliminated 
when two NF baselines (intermediate and long)  were combined \footnote{The authors of \cite{lindner} did not 
find intrinsic degeneracies in their simulations by assuming a very optimistic lower cut 
in the momentum of the muon.}. A comparison of the NF and SPL-SB fits shows
that the displacement of the fake solution with respect to the true one is opposite for the two facilities.

\begin{figure}[t]
\begin{center}
\begin{tabular}{ll}
\hskip -0.5cm
\epsfig{file=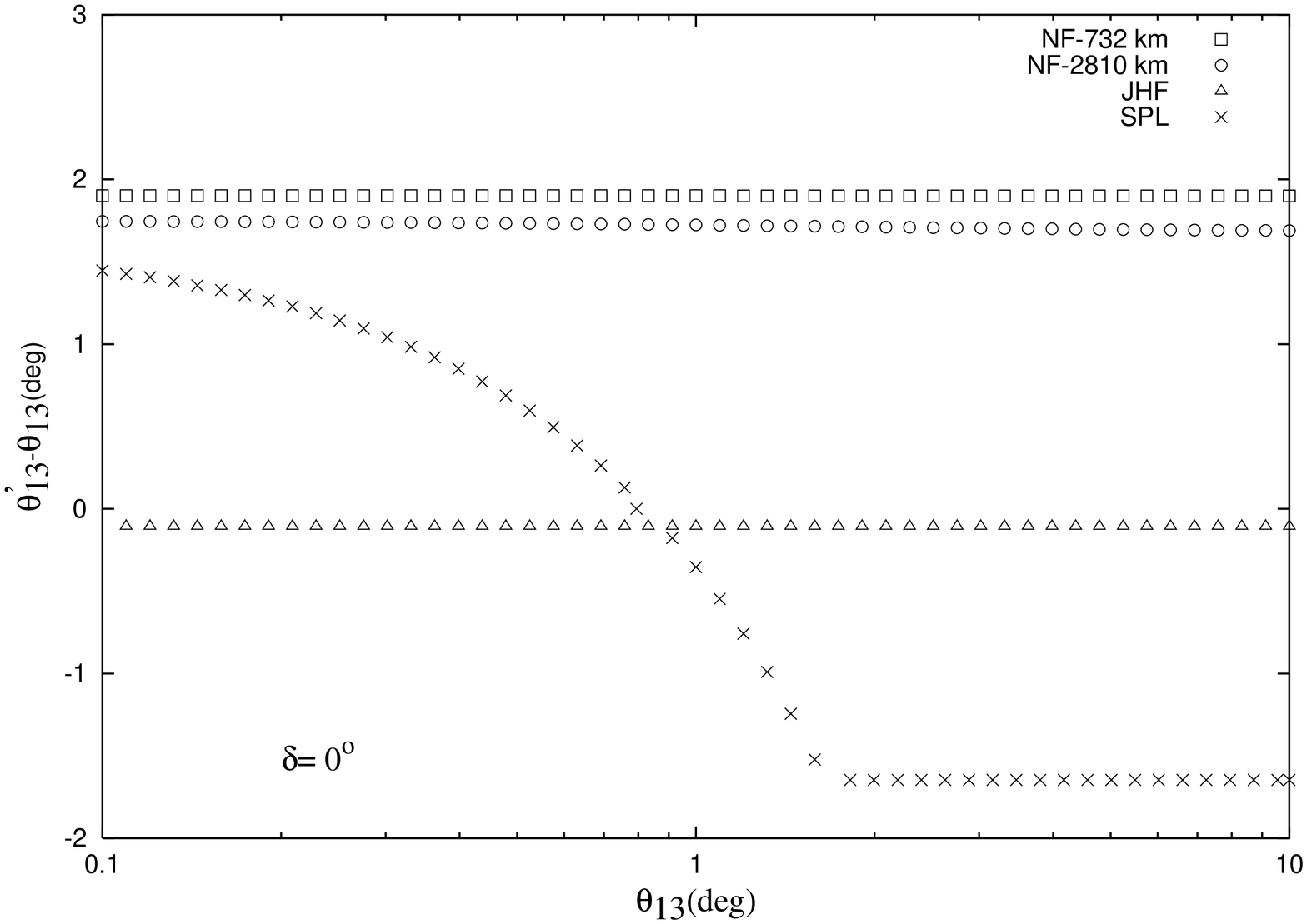, height=8cm,angle=-90} &
\hskip -0.5cm
\epsfig{file=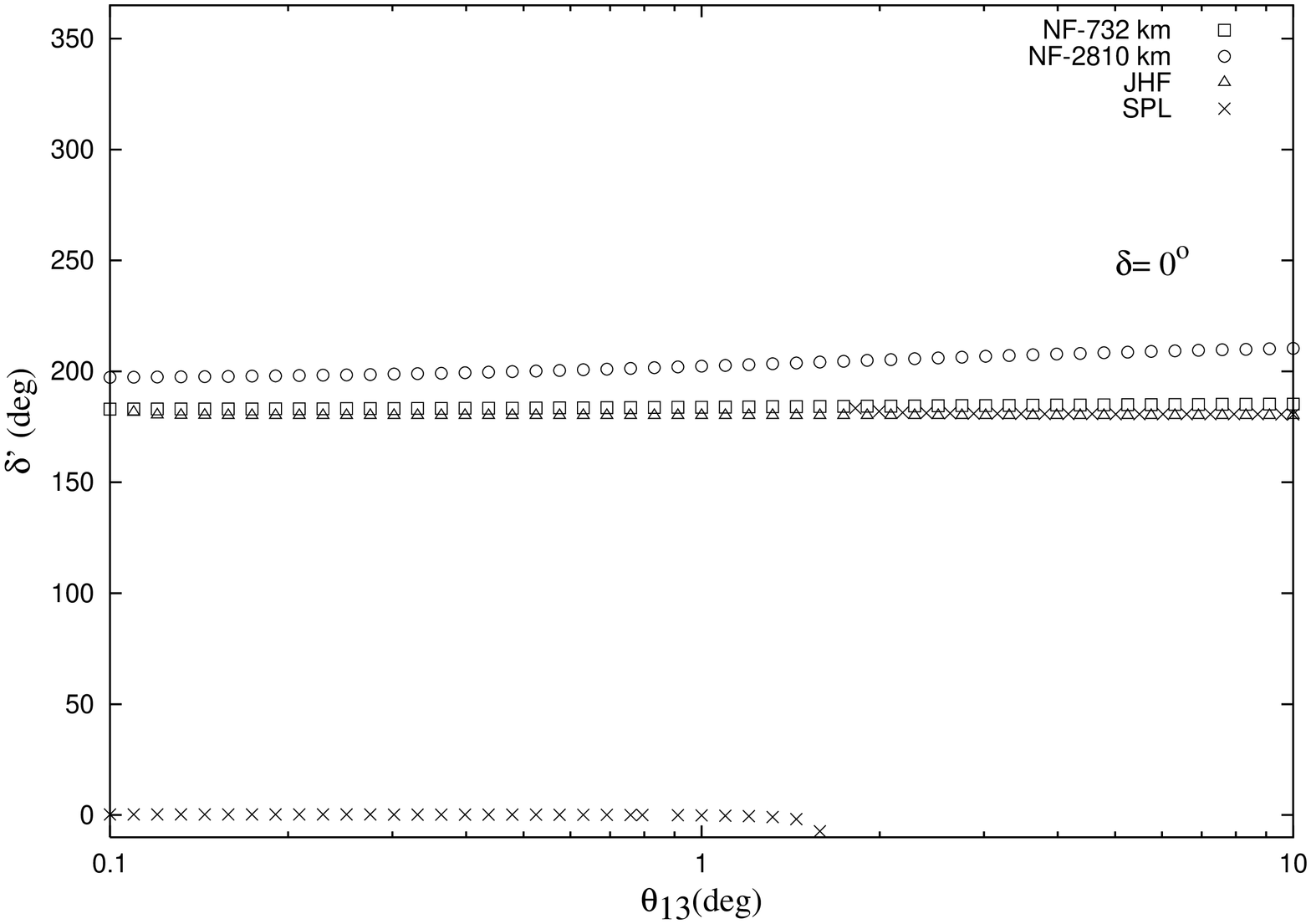, height=8cm,angle=-90} \\
\hskip -0.5cm
\epsfig{file=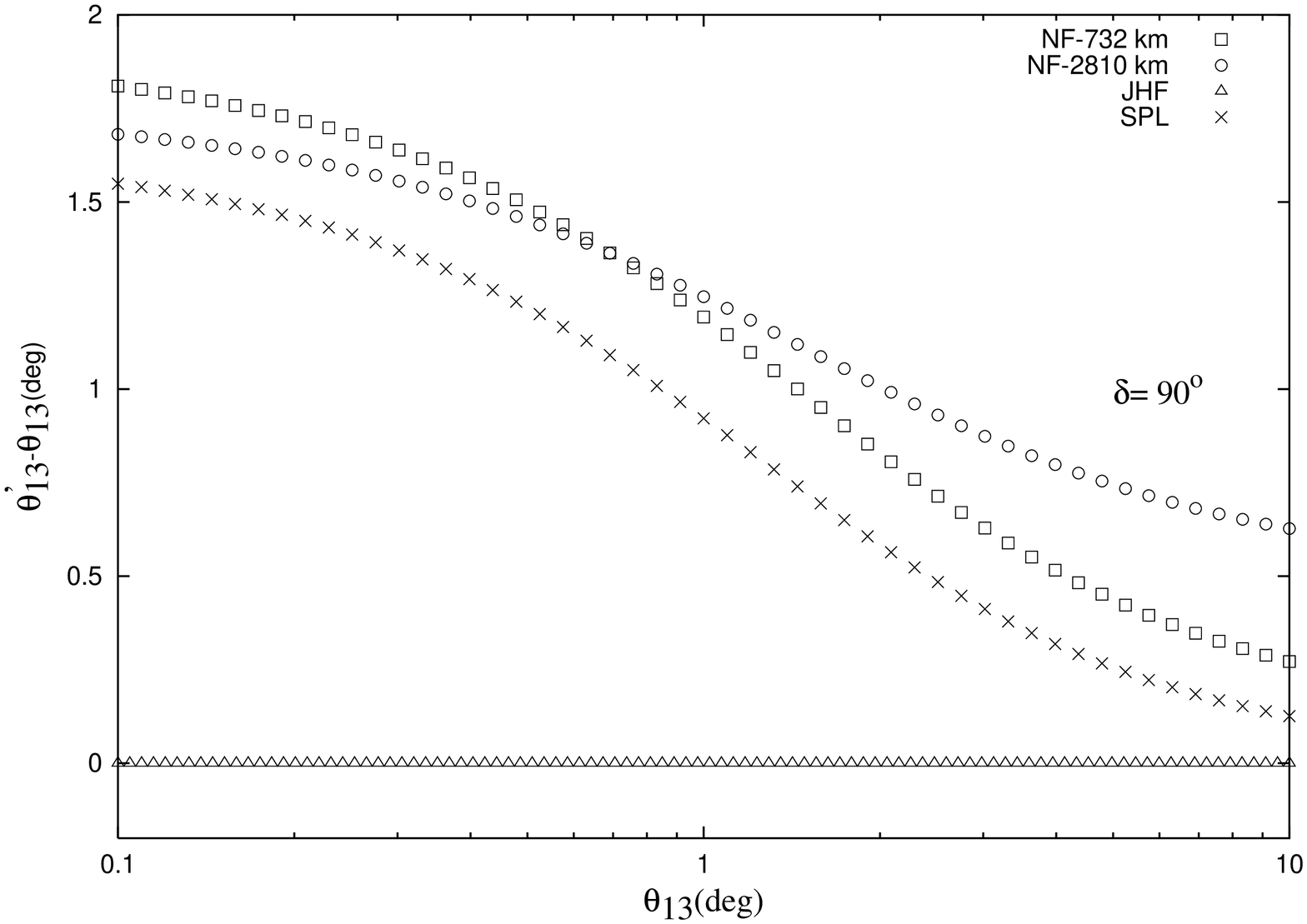, height=8cm,angle=-90} &
\hskip -0.5cm
\epsfig{file=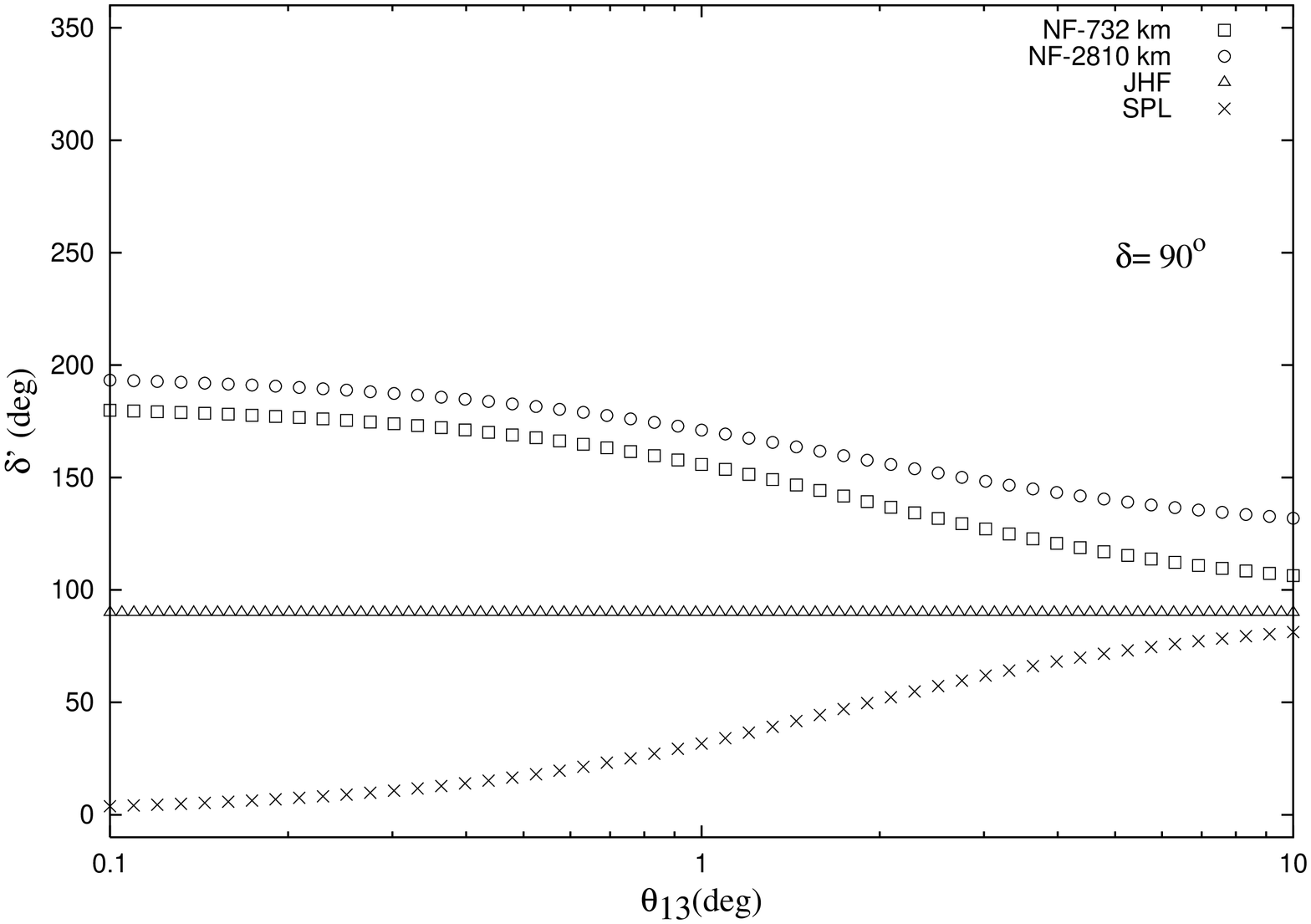, height=8cm,angle=-90} \\
\end{tabular}
\end{center}
\caption{\it{$\tetaotp-\tetaot$ (left) and $\delta'$ (right) versus $\tetaot$, for the intrinsic fake solution, 
for fixed values of $\delta = 0^\circ$(up) and $\delta=90^\circ$ (down).}}
\label{intri1}
\end{figure}

In order to understand the intermediate region between the solar and atmospheric regimes, as well as the influence 
of matter effects, we have determined numerically all 
the possible physical solutions to 
eqs.~(\ref{equalburguet}), using the approximate formulae for the probabilities
including matter effects \cite{golden}. $L$ and $E$ are fixed to the average values for the different facilities. The results for the shift $\tetaotp-\tetaot$ and $\delta'$ are shown 
in Fig.~\ref{intri1} as a function of $\tetaot$, for two values of $\delta=0^\circ,90^\circ$ and for the 
different experimental setups. In the whole range of parameters we find two solutions, as expected by 
periodicity in $\delta$, since one 
solution is warranted: the true one.

\begin{figure}[t]
\begin{center}
\epsfig{file=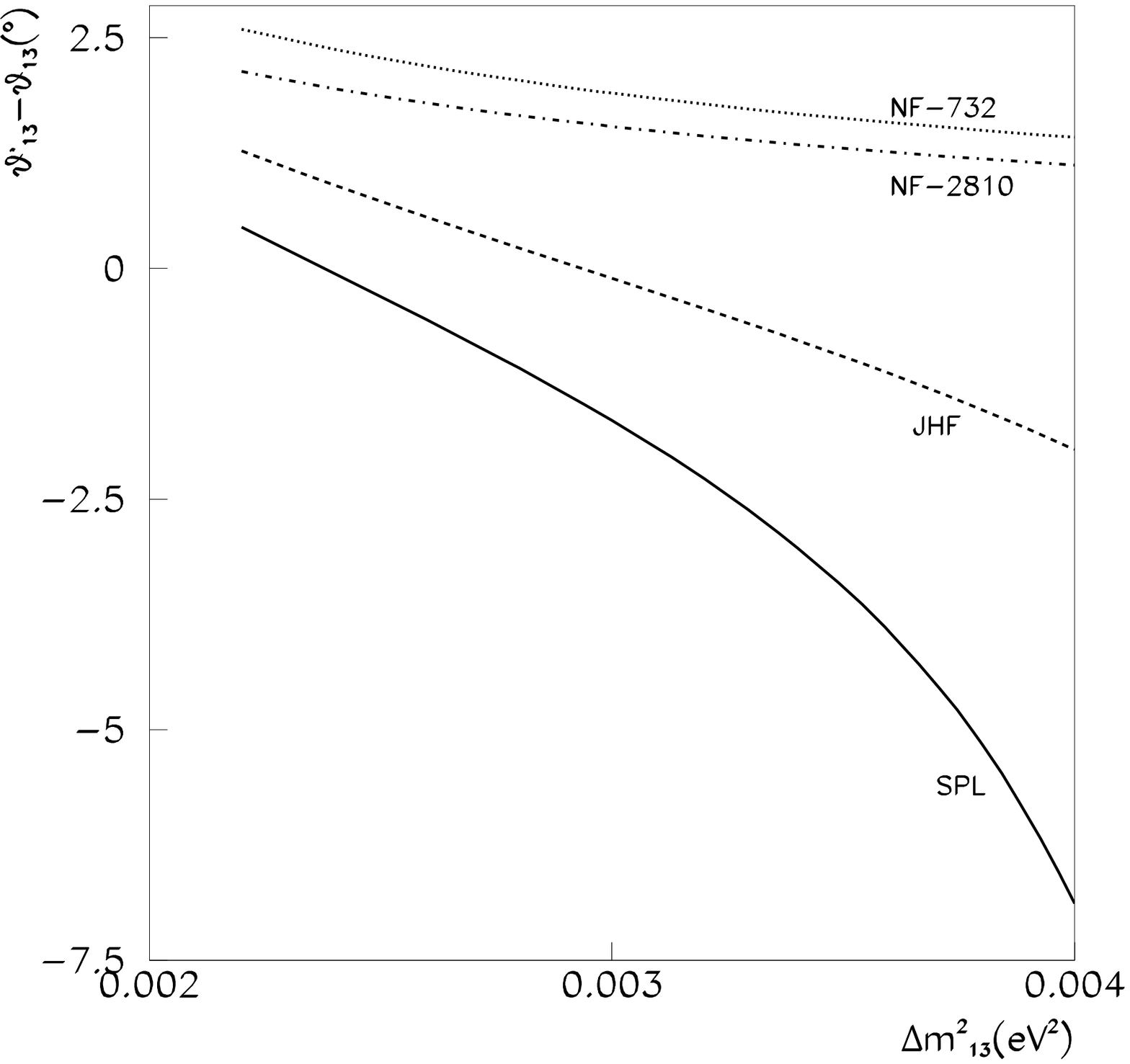, width=8.5cm} \\
\end{center}
\caption{
\it{$\tetaotp-\tetaot$ versus $|\Delta m^2_{13}|$ for the intrinsic fake solution in the atmospheric regime and $\delta=0^\circ$. }}
\label{super8-1}
\end{figure}

The most important point to note in eqs.~(\ref{atmburguet}) and (\ref{solarburguet}) and 
in Figs.~\ref{intri1} is that the position (measured in 
$\tetaotp -\tetaot$ or $\delta '$) of the degenerate solution 
is very different in the NF, the SPL-SB and 
JHF\footnote{For the JHF scenario the shift in $\tetaot$ is minimal because for the reference 
parameters $\enu = 0.7$ GeV and $L= 295$ km,  $\cot\left(\atmos\right)\simeq 0$.} setups.  
As a result, we expect that any combination of the results of two of these
three facilities could in principle exclude the fake solutions. 
The $\tetaotp -\tetaot$ of the fake solution depends strongly on the baseline
and the neutrino energy through the ratio $L/E$,  
  so the combination of the results
of two experiments with a different value for this ratio should be able 
to resolve these degeneracies, within their range of sensitivity. Even more important is 
that, for small $\tetaot$, $\delta'$ may differ by $180^\circ$ if the two facilities have opposite 
sign for  $\cot\left(\atmos\right)$, see eqs.~(\ref{solarburguet}). For the NF setups, this sign is clearly 
positive, since the measurement of CP violation requires, because of the large matter effects, a baseline considerably shorter than that corresponding to the maximum of the
atmospheric oscillation (in vacuum), where the cotangent changes sign. In the SB scenario on the other
hand, because of the smaller $\langle E\rangle$, matter effects are small at
the maximum of the atmospheric oscillation, which then becomes the optimal 
baseline for CP violation studies. It is then not very difficult to ensure that $\cot\left(\atmos\right)$ be dominantly negative in this case\footnote{Note however that the neutrino beams are generally broad 
so it is necessary for this argument to hold that most of the events
have a parent neutrino energy giving the appropiate sign. The results of the fits indicate that this is the case in the two facilities (NF and SPL-SB) 
that are considered in detail. 
}, which results in an optimal complementarity of the two facilities in resolving degeneracies. 

Clearly the
position of the fake solution is very sensitive to the atmospheric $|\Delta m^2_{13}|$. In matter 
we expect a milder dependence, especially if matter effects become dominant. 
In Fig.~\ref{super8-1}
we show the separation in $\tetaot$ of the intrinsic degenerate solution at $\delta = 0^\circ$ in the atmospheric regime  
as a function of $|\Delta m^2_{13}|$. Although in general the separation becomes smaller for 
smaller $|\Delta m^2_{13}|$, it is 
sizeable in the whole allowed range.  The 
relative difference between the results for the NF facility and the SPL superbeam
option is always largest, although the differences between the two superbeams
and that between the NF and JHF are also very large. Note also that the sign 
of $\tetaot^{'} - \theta_{13}$, which is related to that of $\cot\left(\atmos\right)$, is positive in all the domain 
for the NF baselines and negative in most of the domain for 
SPL-SB scenario, which implies that the difference in $\delta'$ between the two facilities 
is $180^\circ$ for small $\theta_{13}$. For JHF, it is negative only for 
$|\Delta m^2_{13}| \geq 0.003$ eV$^2$.  

Concerning the dependence on the solar parameters, it enters only through the 
combination $\sin 2\theta_{12} \,. \sol$. In general $\theta_{13}' -\theta_{13}$ is linear in this quantity, so degenerate 
solutions become closer with smaller $\Delta m^2_{12}$ and also closer to the true solution. Note however that 
$\delta'$ in the solar regime does not depend on the solar parameters and that it differs by $180^\circ$ in the two facilities,
and this separation will remain when $\Delta m_{12}^2$ is lowered.

\begin{figure}[t]
\begin{center}
\mbox{
\epsfig{file=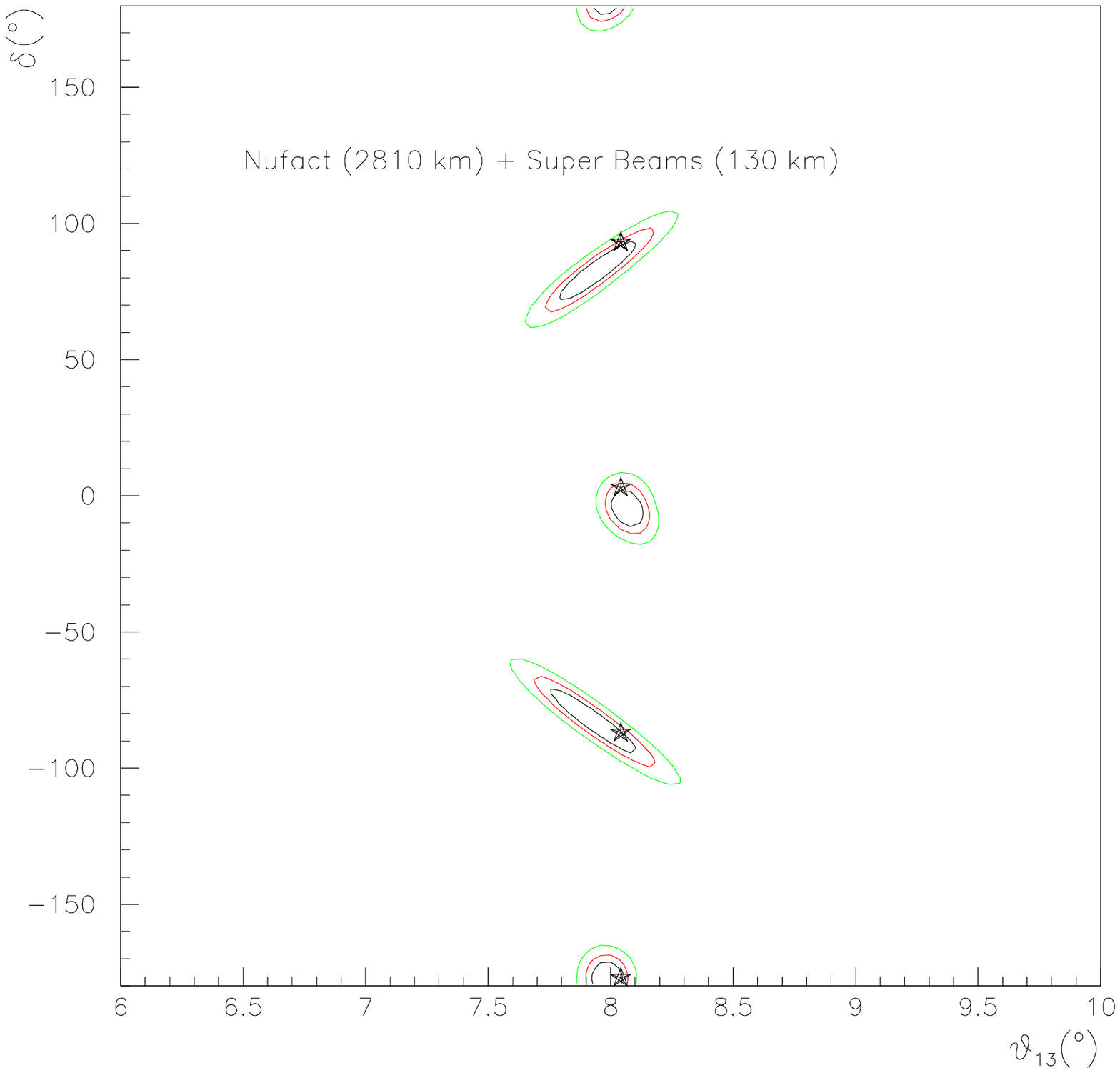,width=8cm}
\epsfig{file=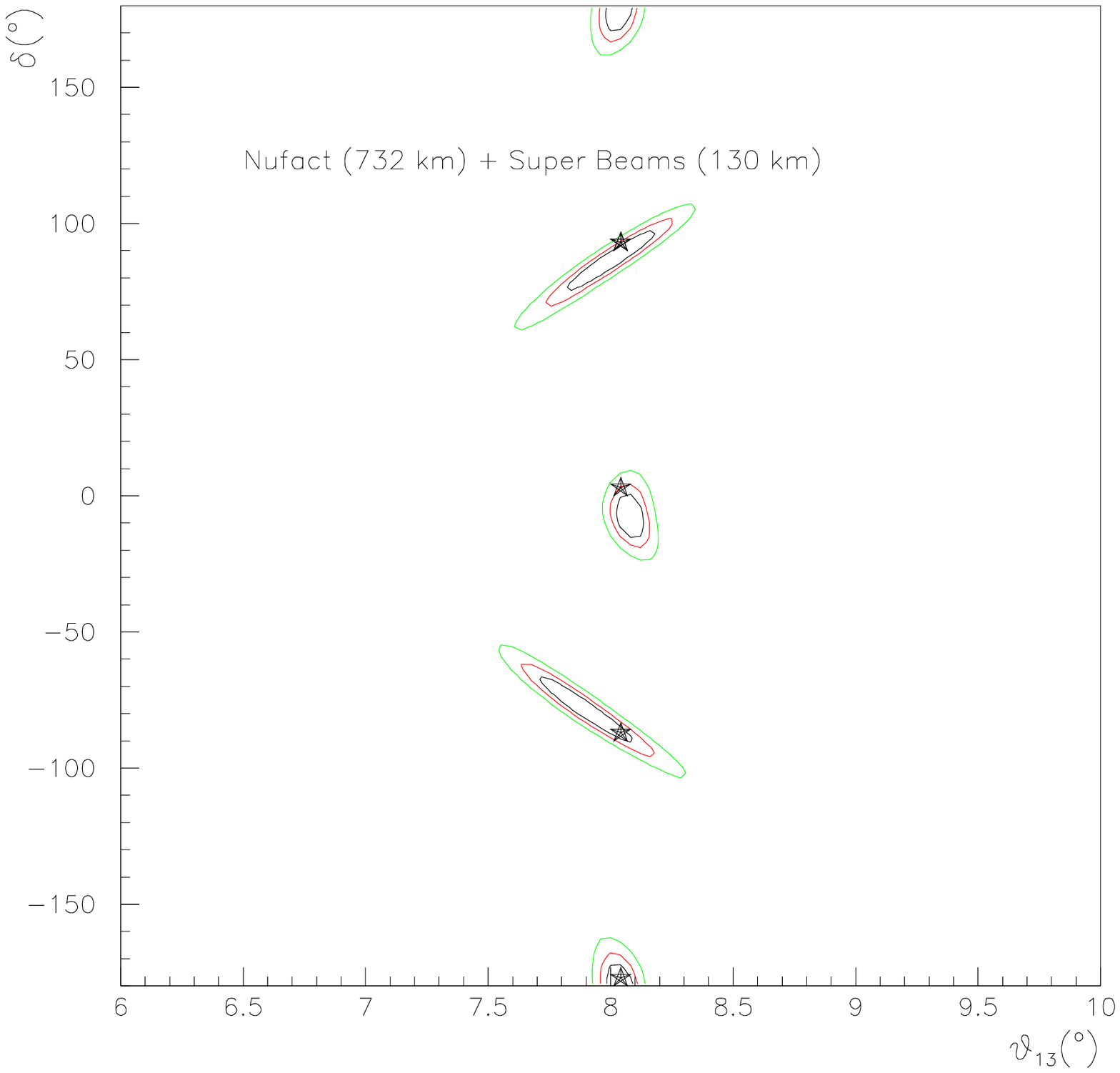,width=8cm}
}
$\tetaot=8^\circ$ 
\\
\mbox{
\epsfig{file=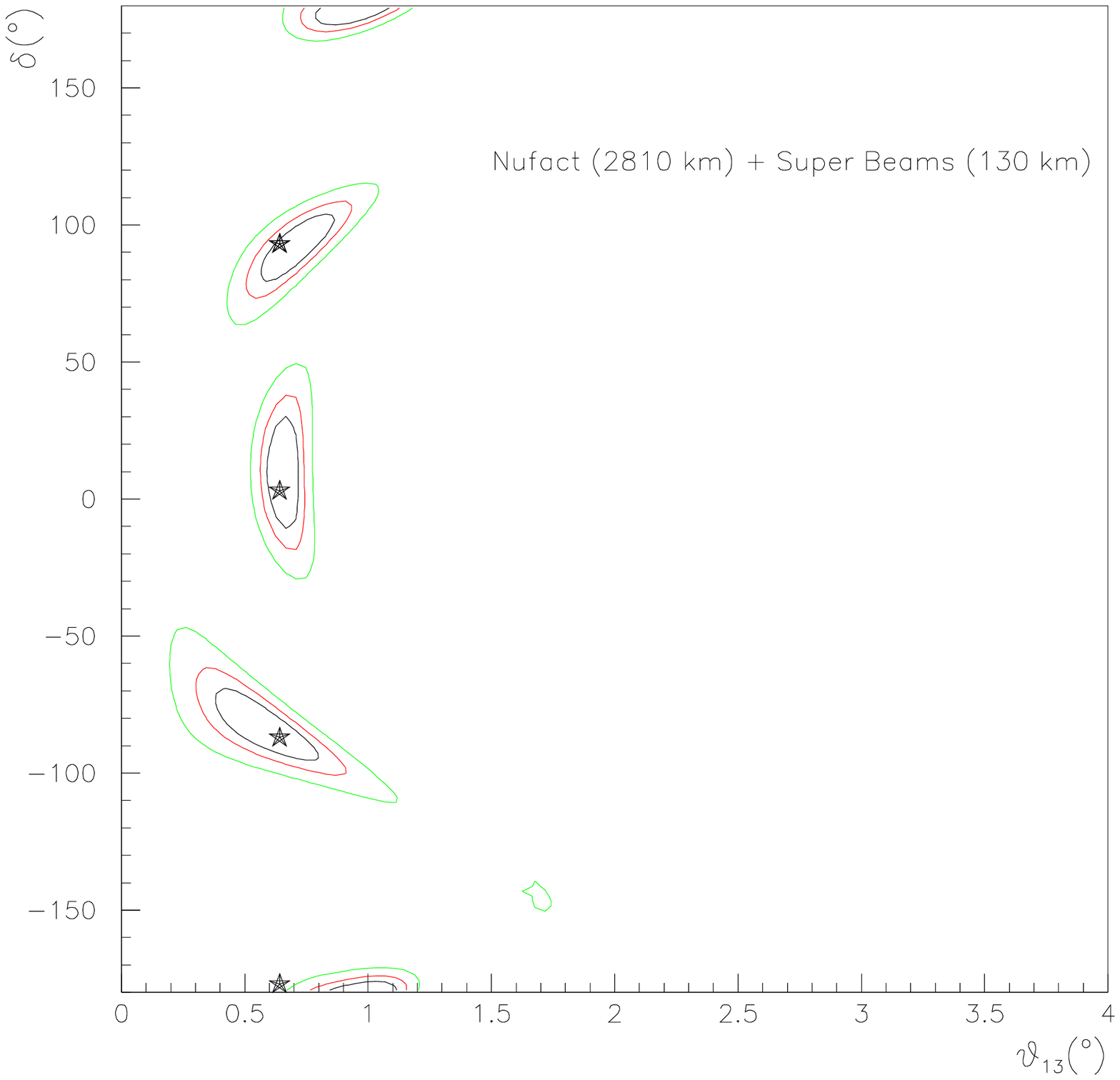,width=8cm} 
\epsfig{file=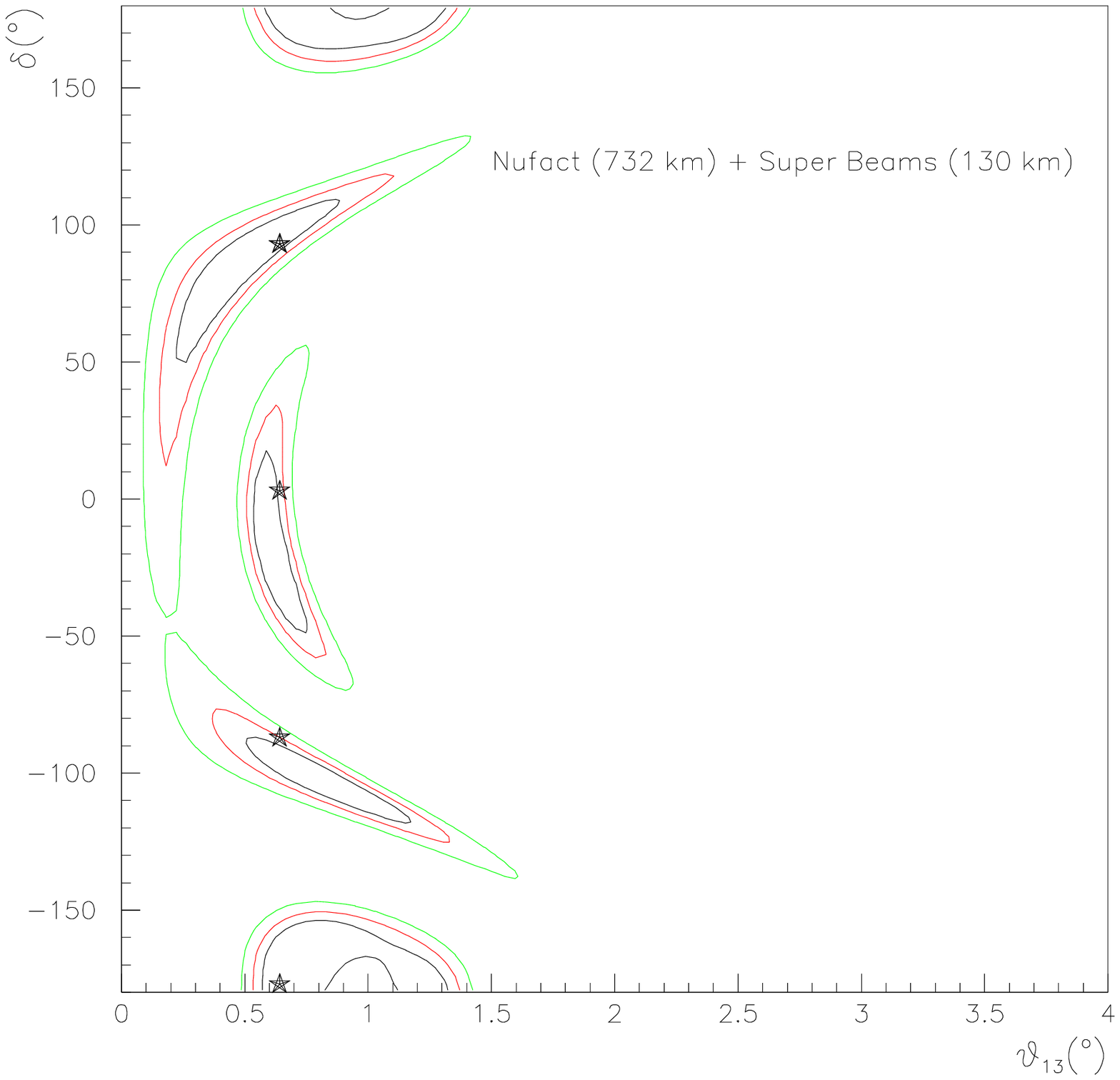,width=8cm}  
}    
$\tetaot=0.6^\circ$
\end{center}
\caption{\it{Fits  
combining the results from the SPL-SB facility and a NF 
baseline at $L= 2810$ km (left)  or $L=732$ km (right).
The true values illustrated correspond to 
 $\delta=- 90^\circ, 0^\circ, 90^\circ, 180^\circ$ and $\tetaot=8^\circ$ (top)
 or $\tetaot=0.6^\circ$ (bottom).
Notice that the fake intrinsic solutions have completely 
disappeared in the combinations.}} 
\label{2810sb}
\end{figure}

 Turning to the variation of the solar parameters while in the atmospheric regime, we will now argue that, if the two facilities that are combined have opposite 
sign($\cot \atmos$), the effect of lowering $\Delta m^2_{12}$ is not 
dramatic either in the resolution of degeneracies.  The statistical  
error on the measurement of $\theta_{13}$ and $\delta$  is  mainly 
independent of the solar parameters (it is dominated by the atmospheric term), which means that at some point when 
$\Delta m^2_{12}$ is lowered, the 
degenerate solutions of the two facilities will merge, since the error 
remains constant while the separation of the solutions gets smaller. However, because of the opposite  
sign of $\tetaot^{'}-\theta_{13}$, the solutions of the two facilities will merge only when they merge with the 
true solution in $\theta_{13}$. 
If this happens, it would therefore not bias the measurement of $\theta_{13}$ 
and $\delta$.

 We have performed a detailed combined analysis of the NF results \cite{golden, burguet} and those from 
the SPL-SB \cite{sb} facility. 
 The 
combination with the optimal NF baseline $L=2810$ km is sufficient to get rid 
of all the fake solutions, as shown in Figs.~\ref{2810sb} (left). Note that indeed the disappearance of the 
fake solutions takes place even in the solar regime!

One very interesting exercise is to reconsider the combination of the SPL-SB and the NF results 
at a shorter baseline of $L=732$ km. As 
explained in refs. \cite{golden, burguet}, the degenerate solution is not so relevant to this
NF baseline when considered alone, because there the sensitivity to CP violation is so poor that 
there exists a continuum of almost degenerate 
solutions, which makes the determination of $\delta$ impossible with the
wrong-sign muon signals. The combination of the results from this NF baseline with 
those from the SPL-SB facility is summarized in the tantalizing plots in Figs.~\ref{2810sb} (right).
Not only do the fake solutions corresponding to the intrinsic 
degeneracies in the superbeam disappear, but the accuracy in the 
determination of the true solution becomes competitive with  
that obtained in the combination with the optimal baseline for large
values of $\tetaot$. At small values of $\tetaot$ the latter still gives better results, as expected.

\section{ ${\rm sign}(\Delta m^2_{13})$ degeneracy} 

 Assume nature has chosen a given sign for $\Delta m^{2}_{13}$, while the data analysis 
is performed assuming the opposite sign. Let us call 
$P^{'}_{\nu_ e\nu_\mu ( \bar \nu_e \bar \nu_\mu ) }(\theta_{13},\delta)$ the 
oscillation probability with the sign of $\Delta m^2_{13}$ reversed. 
 We may then get new fake solutions ($\tetaot^{'} , \delta^{'}$), at fixed $E_\nu$ and $L$, if the equations
\vspace{-0.75cm}
\begin{center}
\begin{equation}
\left.\matrix{
P^{'}_{\nu_e \nu_\mu} (\theta^{'}_{13}, \delta^{'}) = P_{\nu_e \nu_\mu}
(\theta_{13}, \delta)\nonumber \cr 
P^{'}_{\bar \nu_e \bar \nu_\mu} (\theta^{'}_{13}, \delta^{'}) = P_{\bar \nu_e
\bar \nu_\mu} (\theta_{13}, \delta)}
\right \}
\label{nonequalsign}
\end{equation}
\end{center}
have solutions in the allowed physical range. 
\begin{figure}[t]
\begin{center}
\begin{tabular}{ll}
\hskip -0.5cm
\epsfig{file=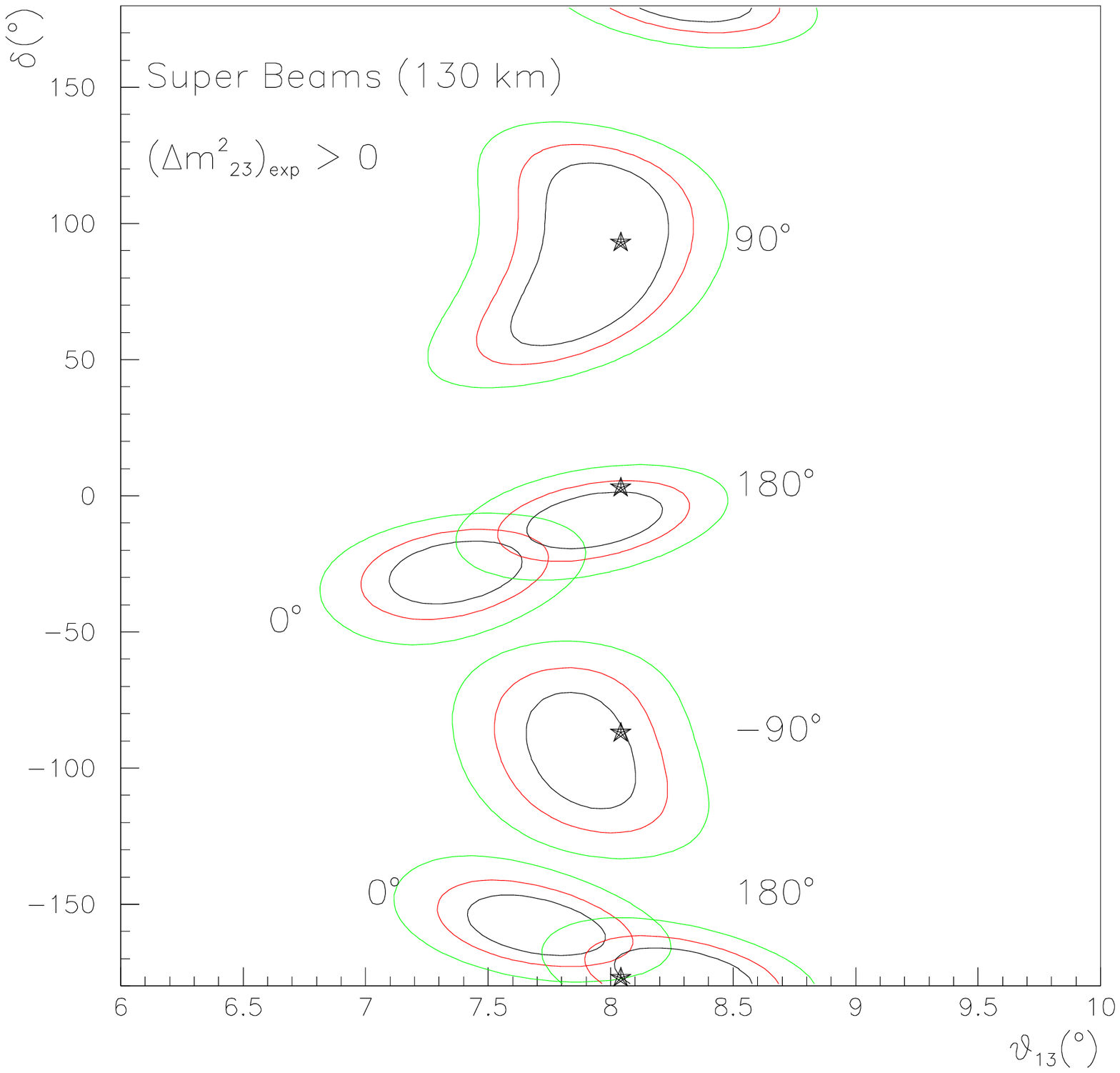, width=8cm} &
\hskip -0.5cm
\epsfig{file=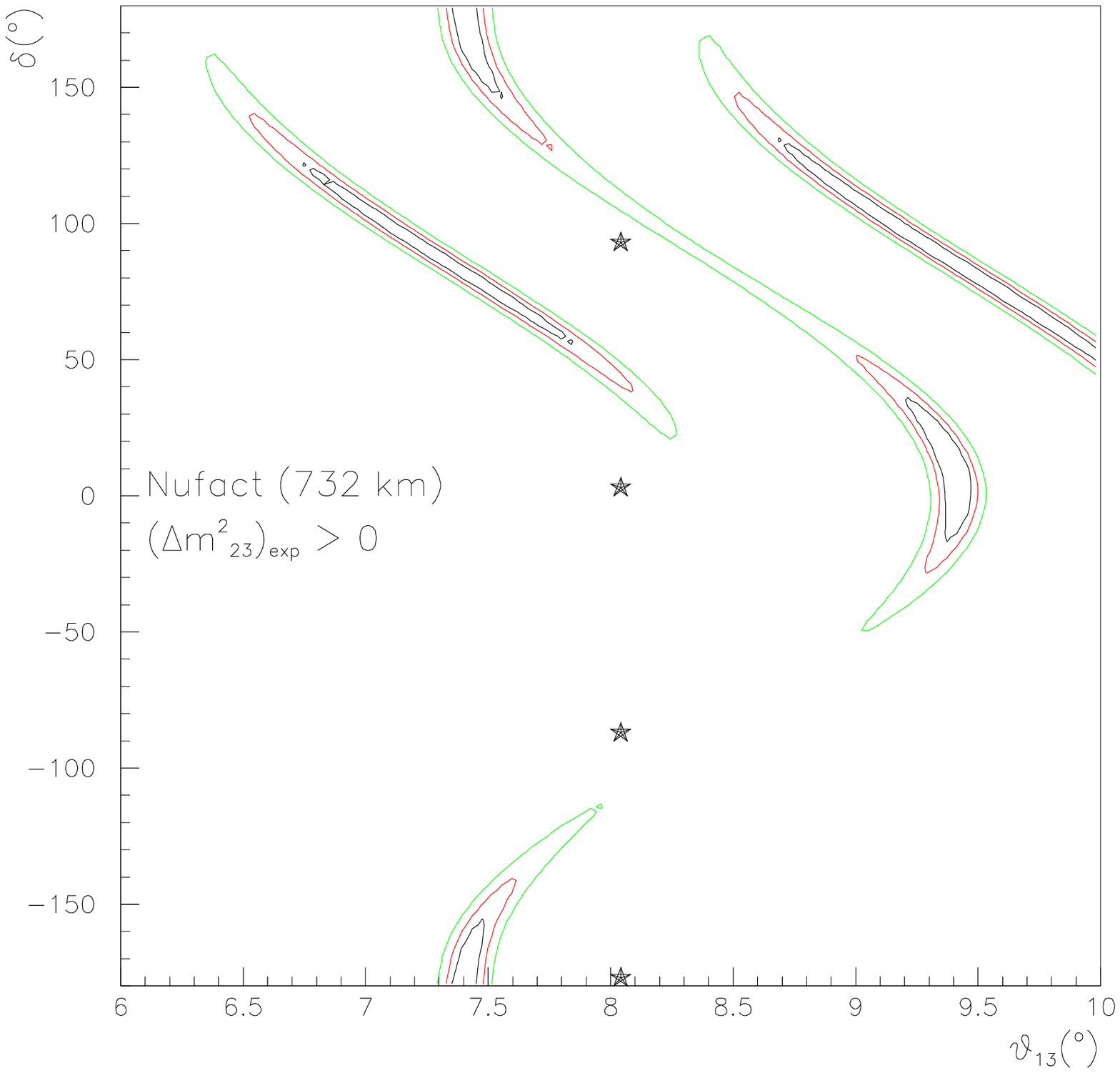, width=8cm} \\
\end{tabular}
\end{center}
\caption{\it{Fits for central 
values $\tetaot=8^\circ$ and  
 $\delta=- 90^\circ, 0^\circ, 90^\circ, 180^\circ$ for the SPL-SB (left) and 
NF at $L=732$ km (right). Nature's
sign for $\Delta m_{23}^2$ is assumed to be positive, while the fits have been performed
 with the opposite sign.   
All fake solutions disappear when the two sets of data are combined.}} 
\label{732sbatmsgn}
\end{figure}

\begin{figure}[t]
\begin{center}
\begin{tabular}{ll}
\hskip -0.5cm
\epsfig{file=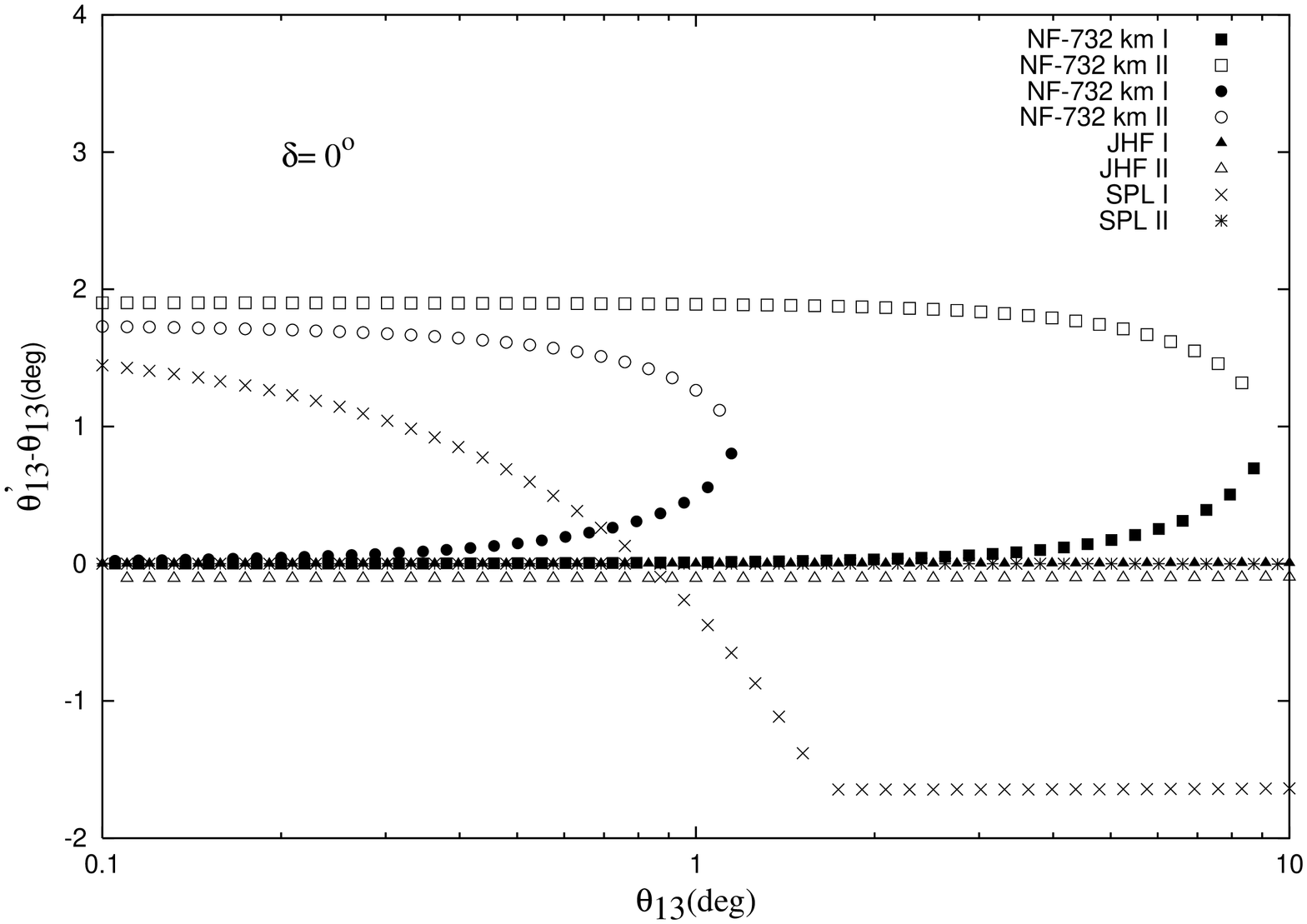, height=8cm,angle=-90} &
\hskip -0.5cm  
\epsfig{file=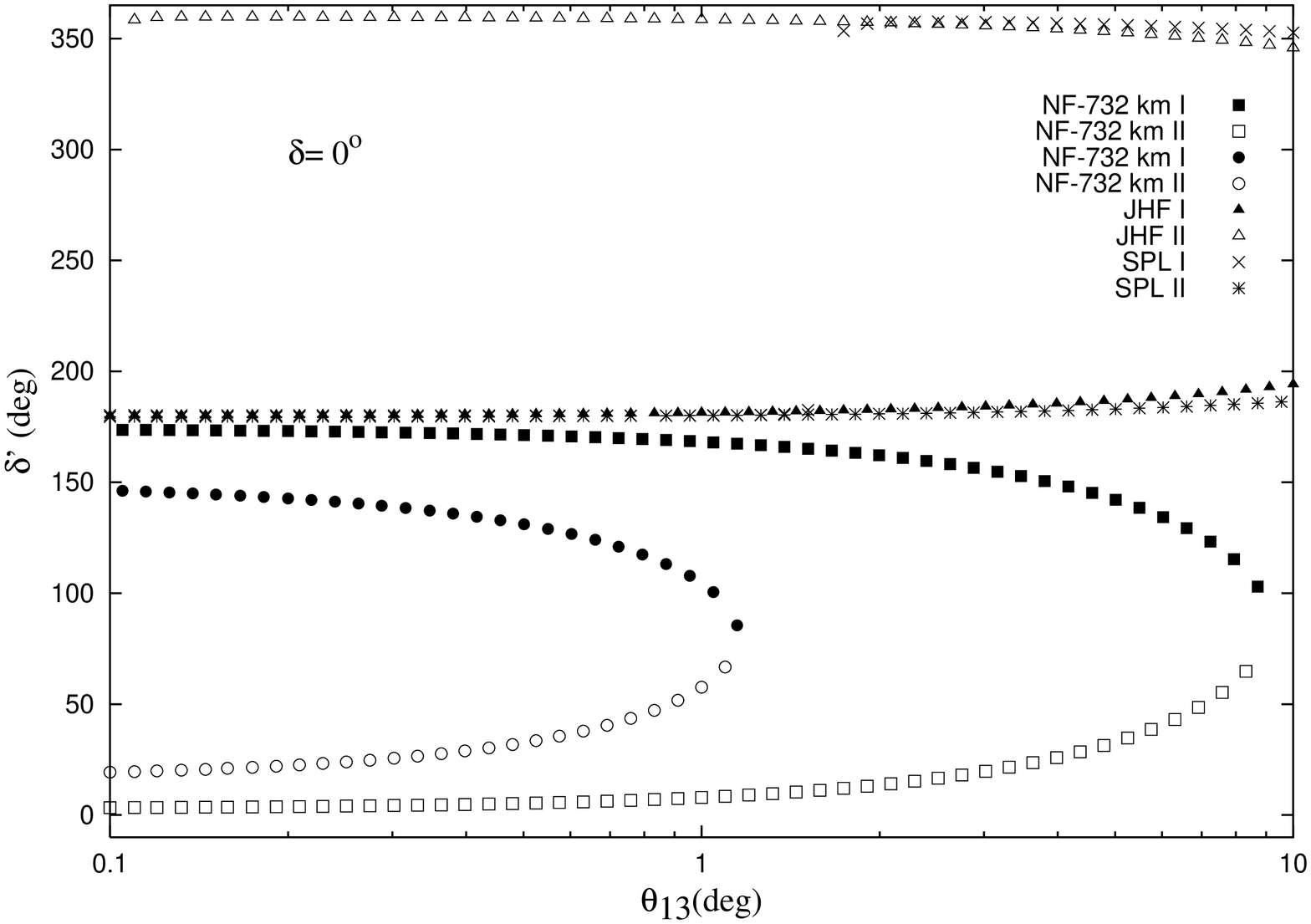, height=8cm,angle=-90} \\
\hskip -0.5cm
\epsfig{file=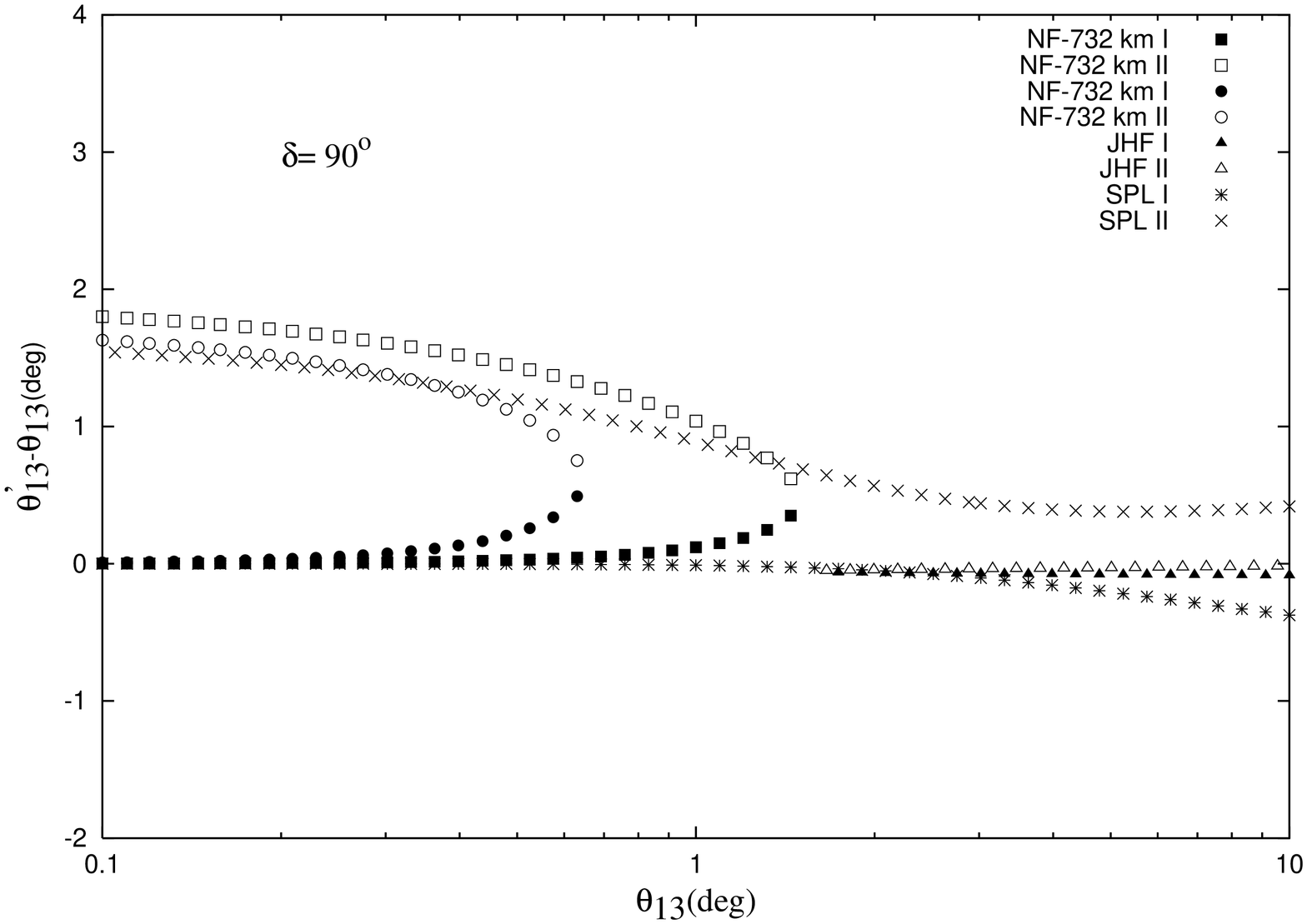, height=8cm,angle=-90} &
\hskip -0.5cm
\epsfig{file=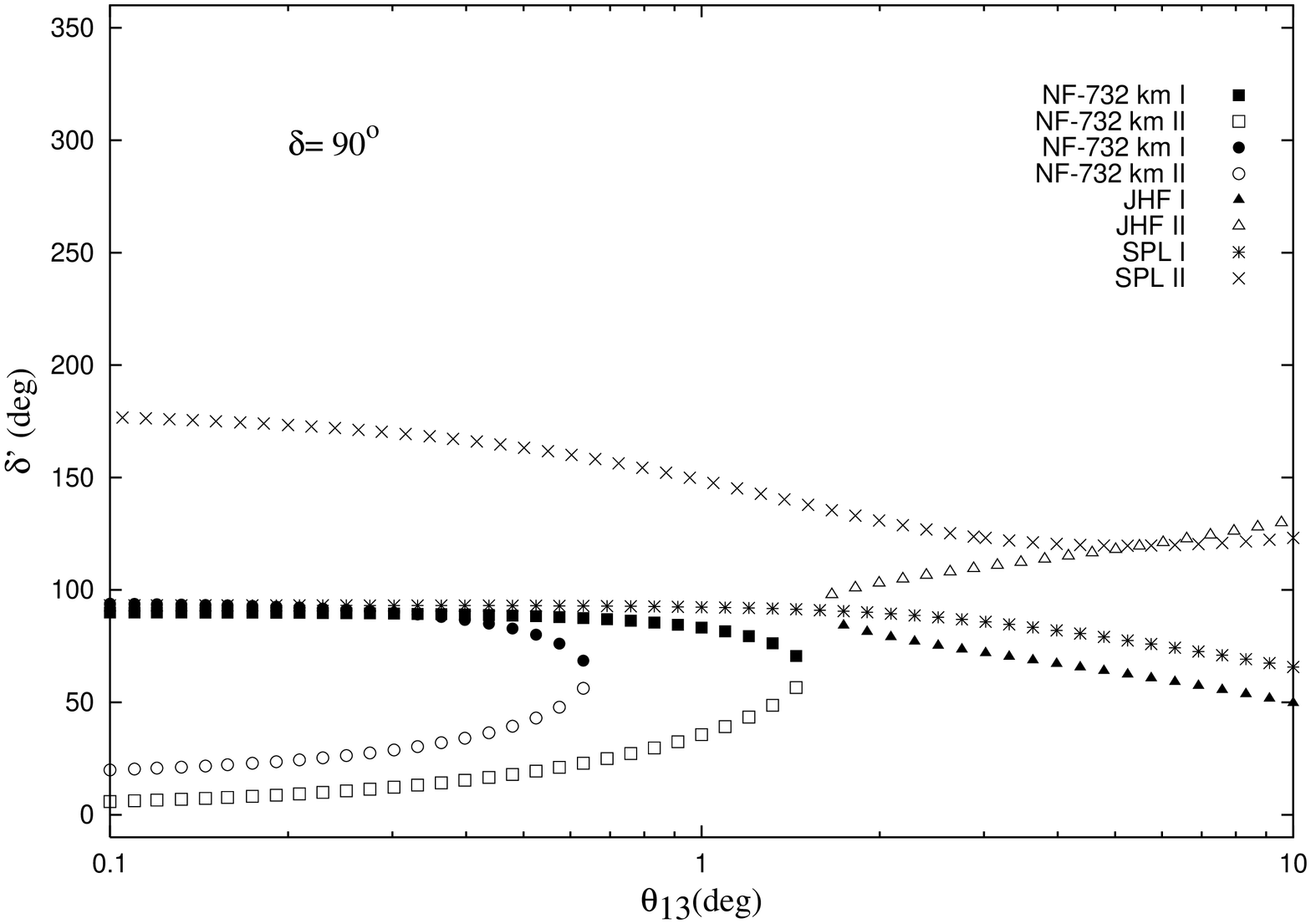, height=8cm,angle=-90} \\
\end{tabular}
\end{center}
\caption{\it{$\tetaotp-\tetaot$ (left) and $\delta'$ (right) for the sign degeneracies as 
functions of $\tetaot$ for fixed values of 
$\delta = 0^\circ$ (up) and $\delta = 90^\circ$ (down). Solutions $I$ and 
$II$ are described in the text. }}
\label{jj2}
\end{figure}

It turns out that there are generically two fake sign solutions to eqs. (\ref{nonequalsign}).
It is very easy to find them in the vacuum approximation, 
as they mirror the two solutions (true and fake) obtained in the analysis of the intrinsic degeneracies. 
It can be seen in eq. (\ref{vacexpand}) that a change in the sign of $\Delta m^2_{13}$ 
can be traded in vacuum by the substitution $\delta \rightarrow \pi - \delta$ \cite{mnp}, implying then 
for eqs.~(\ref{nonequalsign}) 
\bea
P^{'}_{\nu_e \nu_\mu(\bar{\nu}_e \bar{\nu}_\mu)} (\theta^{'}_{13}, \delta^{'})
\simeq 
P_{\nu_e \nu_\mu(\bar{\nu}_e \bar{\nu}_\mu)} (\tetaotp, \pi-\delta') \,,
\eea
in the vacuum approximation. 
 Consequently, the solutions in vacuum can be obtained from those present for the intrinsic case, upon 
the substitution $\delta' \rightarrow \pi - \delta'$. 
One of them mirrors the true (nature) solution and will be called below solution I, given in vacuum by 
\bea
\delta^{'}&\simeq&\pi -\delta,\nonumber\\
\theta^{'}_{13}&\simeq & \theta_{13}.  \; \;
\label{signI}
\eea 
The fact that it is approximately $E$- and $L$-independent suggests that it will be hard to eliminate it by 
exploiting  the $L,E$ dependence of different facilities, as indeed is confirmed by the fits below. 
Fortunately, this fake solution does not interfere 
significantly with the determination of $\tetaot$
or CP-violation (i.e. $\sin \delta$).

The second fake sign solution, which we will call solution II, can be read 
in vacuum  
from eqs.~(\ref{atmburguet}) and (\ref{solarburguet}), 
upon the mentioned $\delta^{'} \rightarrow \pi - \delta^{'}$ exchange. It is strongly $L$- and $E$-dependent.
 Both solutions I and II can be nicely seen in the numerical analysis for the SPL-SB in Fig.~\ref{732sbatmsgn} (left), for $\tetaot=8^\circ$ and positive 
sign$(\Delta m_{13}^2)$.

Matter effects are obviously very important in resolving fake sign 
solutions\cite{sign}: the task should thus be 
easier at large $\tetaot$ and large enough  NF 
baselines, where matter effects are largest\footnote{ We have checked that for the sign degeneracies 
(and only for these), the combined results from a SB facility and  
 a NF baseline  at $L = 7332$ km are competitive or even superior to those 
obtained in the combination with an intermediate NF baseline.}.
In fact it is easy to prove that 
no solutions can remain for large
enough $\tetaot$.
This can be seen in Figs.~\ref{jj2}, which show the fake sign 
solutions as they result from solving numerically eqs.~(\ref{nonequalsign}) (using the approximate 
probabilities with matter effects included \cite{golden}) for the different experiments. 
For small $\tetaot$ the two solutions I and II exist in all cases, while for large $\tetaot$ they degenerate 
and disappear because of matter effects. One should keep in mind, though, that 
even if no fake solution exists, there might be approximate ones that 
will show up in a measurement with finite errors.  

Our fits including realistic background errors and efficiencies confirm  
the above expectations, at each given facility. To be more precise, we have found no fake sign solution
for values of 
$\tetaot > 2^\circ$, when considering just one NF baseline of $L=2810$ km (or longer), while for $2^\circ>\tetaot>1^\circ$
 they do appear but get eliminated when the data are combined with those from the SPL-SB.
At $L=732$ km some fake sign solutions remain close to the present experimental limit for $\tetaot$, as shown in 
Figs.~\ref{732sbatmsgn} (right). It should be noticed that, again, in the combination of these latter data with those from the 
SPL-SB facility, all fake sign solutions disappear for large $\tetaot \geq 4^\circ$, and the sign 
of $\Delta m_{13}^2$ could thus be determined from it.

\begin{figure}[t]
\begin{center}
\begin{tabular}{ll}
\hskip -0.5cm
\epsfig{file=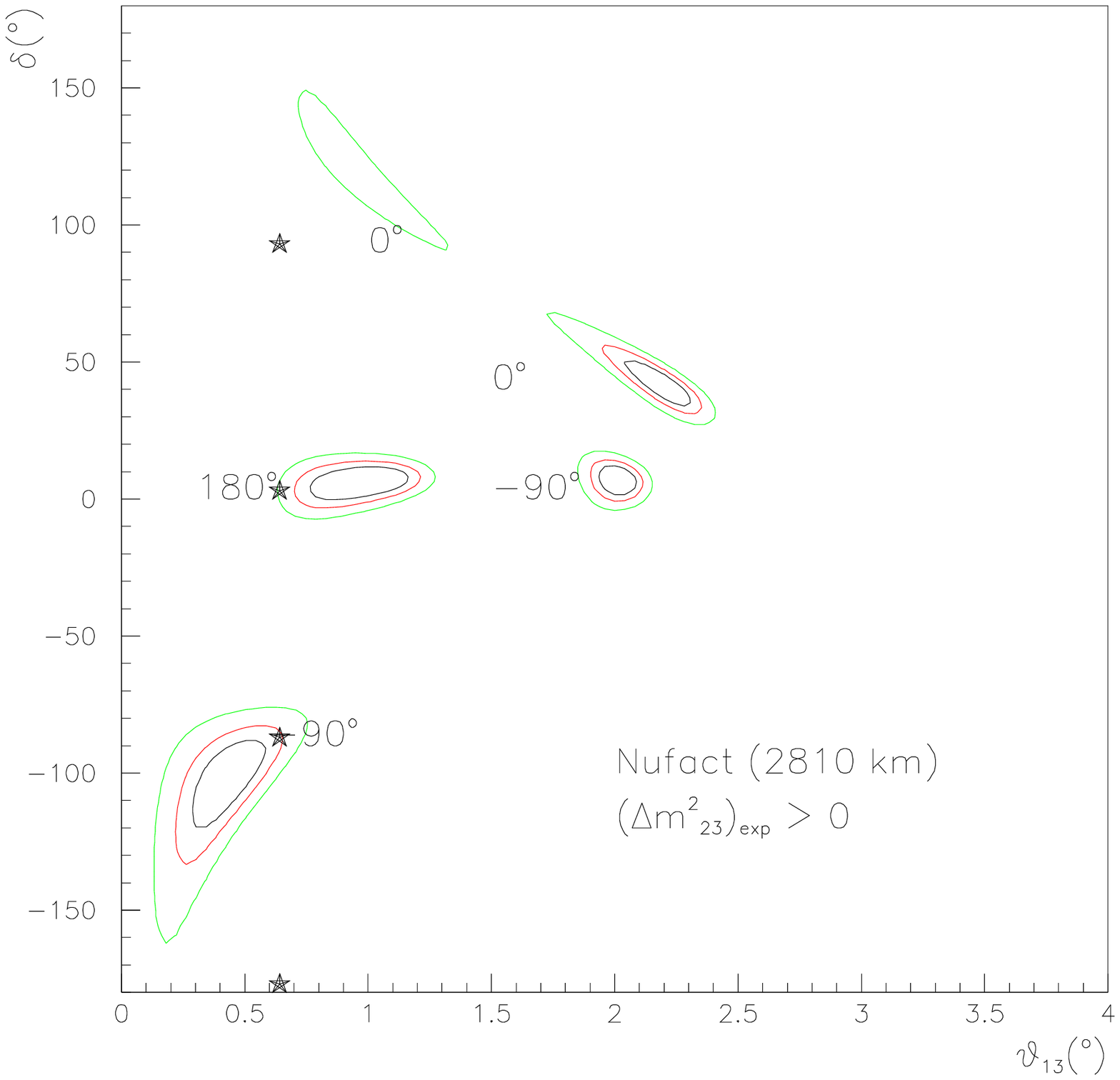, width=8cm} &
\hskip -0.5cm
\epsfig{file=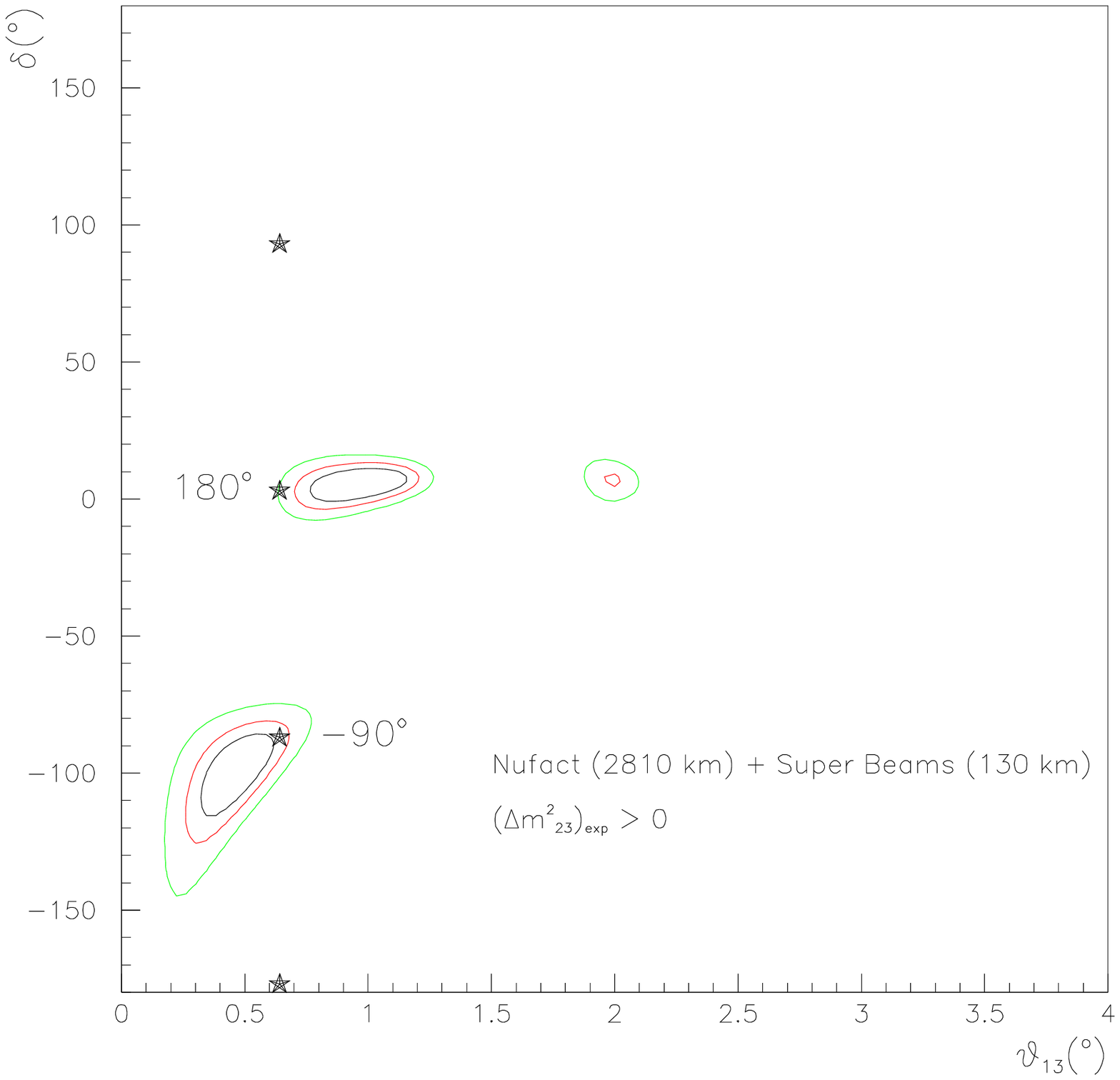, width=8cm} \\
\end{tabular}
\end{center}
\caption{\it{Fits resulting in fake sign solutions, for central 
values $\tetaot=0.6^\circ$ and  
 $\delta=- 90^\circ, 0^\circ, 90^\circ, or 180^\circ$. The nature 
sign for $\Delta m_{13}^2$ is positive, while the fits have been performed
 with the opposite sign.  The results from an NF baseline at $L=2810$ km can be appreciated on the 
left, while their combination with data from the SPL-SB facility can be seen on the right.}} 
\label{2810solarsgn}
\end{figure}

Figures~\ref{jj2} also illustrate that solution I is more facility-independent  
than solution II, as argued above. The solutions that survive in the combinations for small $\theta_{13}$ are 
indeed of type I, as shown in 
Figs.~\ref{2810solarsgn} \footnote{In the same exercise, but with the opposite sign of $\Delta m_{13}^2$, which 
leads to larger statistics, solution II disappears completely.}.

In conclusion, in the LMA-SMW regime, the sign of $\Delta m_{13}^2$ can be determined from data at an intermediate or long NF baseline alone for
$\tetaot$ well inside the atmospheric regime. For the larger values of $\tetaot$, the combination of data 
from the SB facility and a $L=732$ km NF baseline also results in no fake sign solutions.

 With lowering $\tetaot$ ($\tetaot > 1^\circ$ for our central parameters),  the sign can still be determined 
through the combination of SB and NF data at the intermediate or long 
distance. 

 Finally, for the range  $\tetaot < 1^\circ$, the sign cannot be determined, but the combination of data from the SB facility 
and an intermediate (or long) NF baseline is still important to reduce the fake solutions to those of type I, which do not
 interfere significantly with the determination of  $\tetaot$ and $\delta$.

Concerning the dependence on the solar parameters, we do not expect that
the conclusions will change very much with lower $\sin 2 \theta_{12} \Delta m^2_{12}$. The argument for solutions of type II parallels that given in
the previous section for the intrinsic fake solution, while the existence
and position of the type I solutions is pretty insensitive to the solar
parameters. 

%%%%%%%%%%%%%%%%%%%%%%%%%%%%%%%%%%%%%%%%%%%%%%%%%%%%%%%%%%%%%%%%%%%%%%%%%%%%%%%
\section{$\tetatt\rightarrow\pi/2 - \tetatt$  degeneracy}
%%%%%%%%%%%%%%%%%%%%%%%%%%%%%%%%%%%%%%%%%%%%%%%%%%%%%%%%%%%%%%%%%%%%

The present atmospheric data indicate that $\tetatt$ is close to maximal, although not necessarily $45^\circ$.
Superkamiokande results \cite{toshito} give
$90\%$CL-allowed parameter regions for $\sin^{2} 2 \theta_{23}>0.88$, translating into 
the allowed range $35^\circ<\theta_{23}<55^\circ$. 
Therefore even if the value of $\sin^2 2 \theta_{23}$ is determined with 
 great accuracy in disappearance measurements, there may remain 
a discrete ambiguity under the interchange $\tetatt \leftrightarrow \pi/2 - \tetatt$. 
If this $\tetatt$ ambiguity is not cleared up by the time of the NF operation, supplementary fake solutions \cite{bargerdeg} may appear 
when extracting $\tetaot$ and $\delta$, when the wrong choice of octant 
is taken for $\theta_{23}$. 
Fake solutions follow from solving the system of equations, for fixed $L$ and $E_\nu$:
\vspace{-0.75cm}
\begin{center}
\begin{equation}
\left.\matrix{
P^{''}_{\nu_e \nu_\mu} (\theta^{'}_{13}, \delta^{'}) = P_{\nu_e \nu_\mu}
(\theta_{13}, \delta)\nonumber \cr 
P^{''}_{\bar \nu_e \bar \nu_\mu} (\theta^{'}_{13}, \delta^{'}) = P_{\bar \nu_e
\bar \nu_\mu} (\theta_{13}, \delta)}
\right \}\;\;,
\label{nonequalth}
\end{equation}
\end{center}
where $P^{''}_{\nu_e \nu_\mu}$ denotes the oscillation probabilities upon the exchange 
  $\tetatt \leftrightarrow \pi/2 - \tetatt$.

It turns out that, within the allowed range for the parameters, there are generically two solutions 
to these equations. 
They should converge towards the true solution and its intrinsic degeneracy, in the 
limit $\tetatt \rightarrow \pi/4$. We will thus denote again solution I that which mirrors nature's choice and 
solution II that which mirrors the intrinsic degeneracy. Because of this parenthood, solution I is {\it a priori} 
expected to present generically less $L$ and $E$ dependence than solution II,
and be thus more difficult to eliminate in the combination.

It is easy and simple to obtain the analytical form of the fake degeneracies in the vacuum approximation, 
in which, from eqs.~(\ref{vacexpand}) we get 
\bea
P^{''}_{\nu_ e\nu_\mu ( \bar \nu_e \bar \nu_\mu ) }(\tetaotp,\delta') & = & 
c_{23}^2 \, \sin^2 2 \tetaotp \, \sin^2 \atmos  + 
s_{23}^2 \, \sin^2 2 \theta_{12} \, \sin^2\left(\sol\right) \nn \\
& + & \tilde J^{'} \, \cos \left ( \deltap \mp \atmos \right ) \;
 \sol \sin\left(\atmos\right) \,\,.
\label{nmmeq} 
\eea

%%%%%%%%%%%%%%%%%%%%%%%%%%%%%%%%%%%%%%%%%%%%%%%%%%%%%%%%%%%%%%%%%%%%%%%%%%%%%%%
\begin{figure}[t]
\begin{center}
\begin{tabular}{ll}
\hskip -0.5cm
\epsfig{file=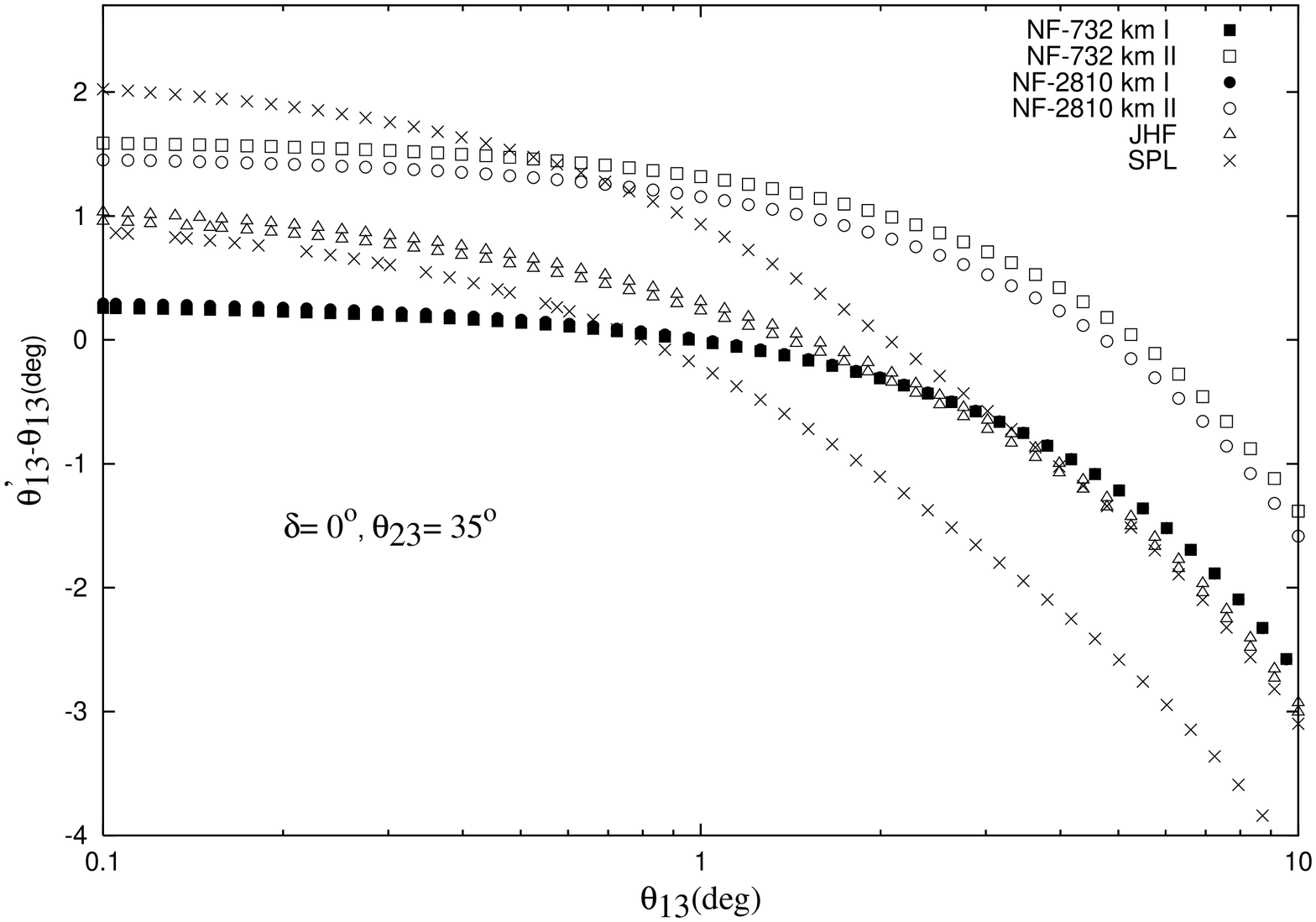, height=8cm,angle=-90} &
\hskip -0.5cm
\epsfig{file=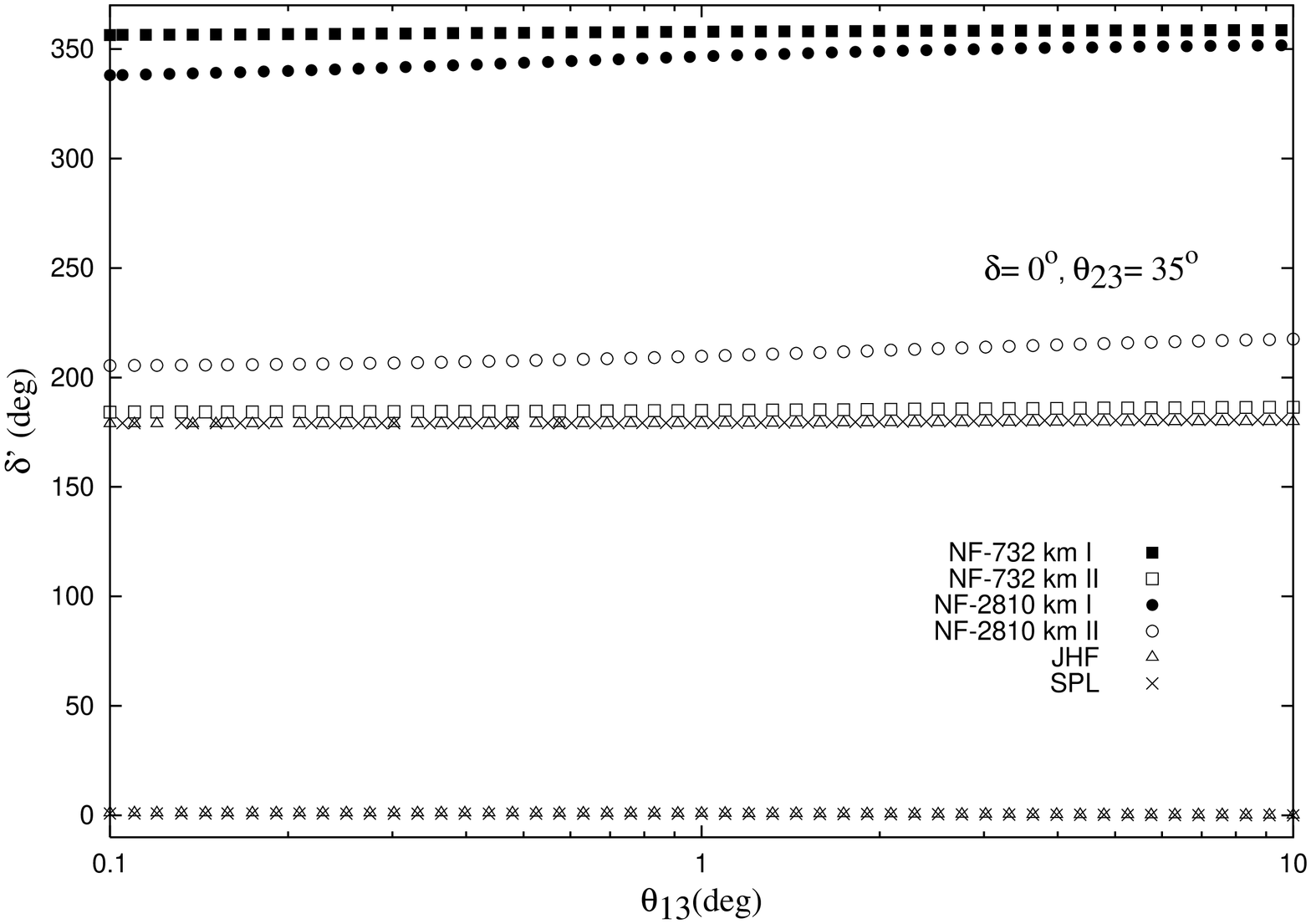, height=8cm,angle=-90} \\
\hskip -0.5cm
\epsfig{file=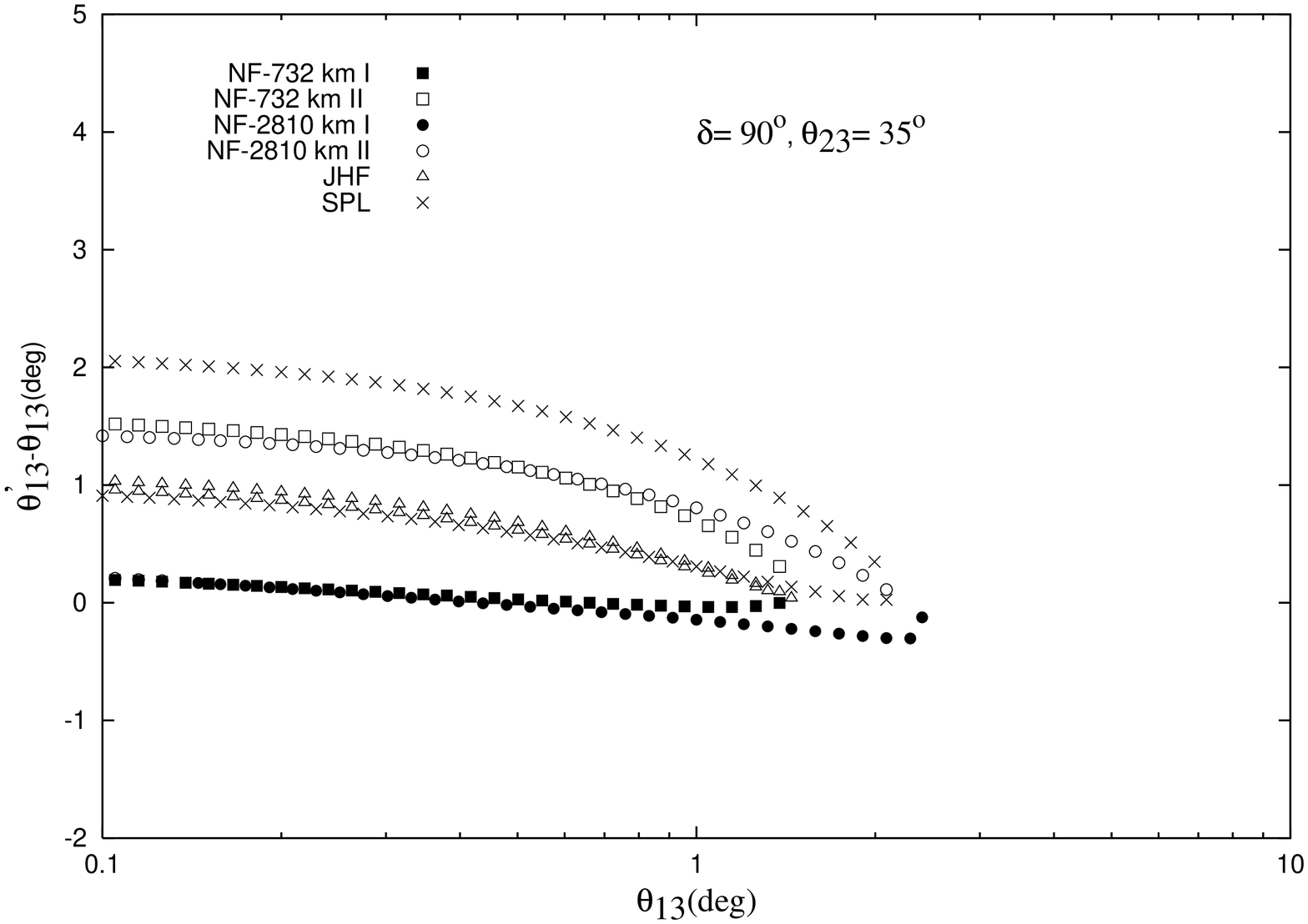, height=8cm,angle=-90} &
\hskip -0.5cm
\epsfig{file=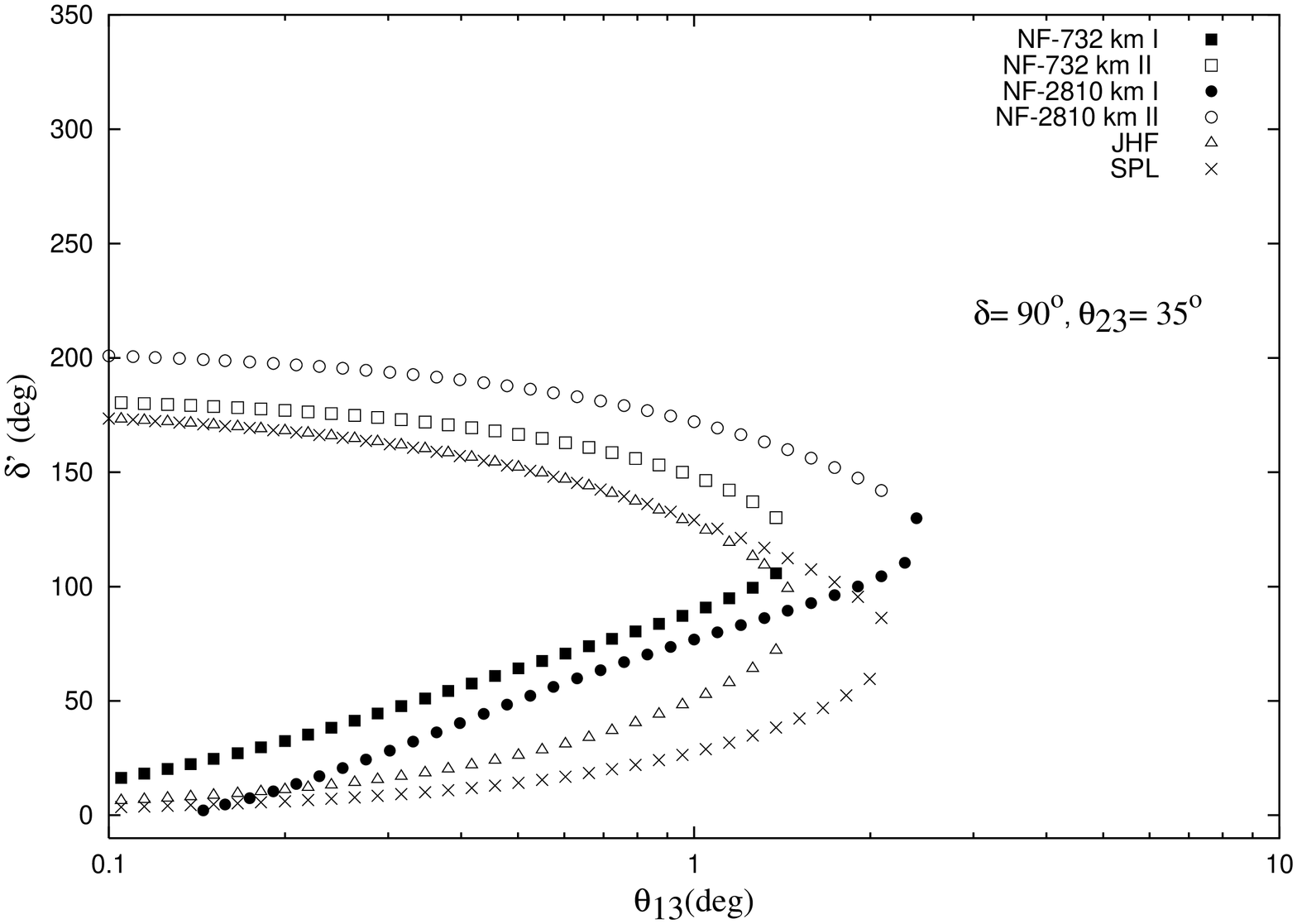, height=8cm,angle=-90} \\
\end{tabular}
\end{center}
\caption{\it{$\tetaotp-\tetaot$ (left) and $\delta'$ (right) for the $\tetatt$ fake solution as functions of $\tetaot$, 
for $\tetatt=35^\circ$, for fixed values of 
$\delta = 0^\circ$(up) and $\delta=90^\circ$ (down).}} 
\label{nmm35}
\end{figure}
\begin{figure}[t]
\begin{center}
\begin{tabular}{ll}
\hskip -0.5cm
\epsfig{file=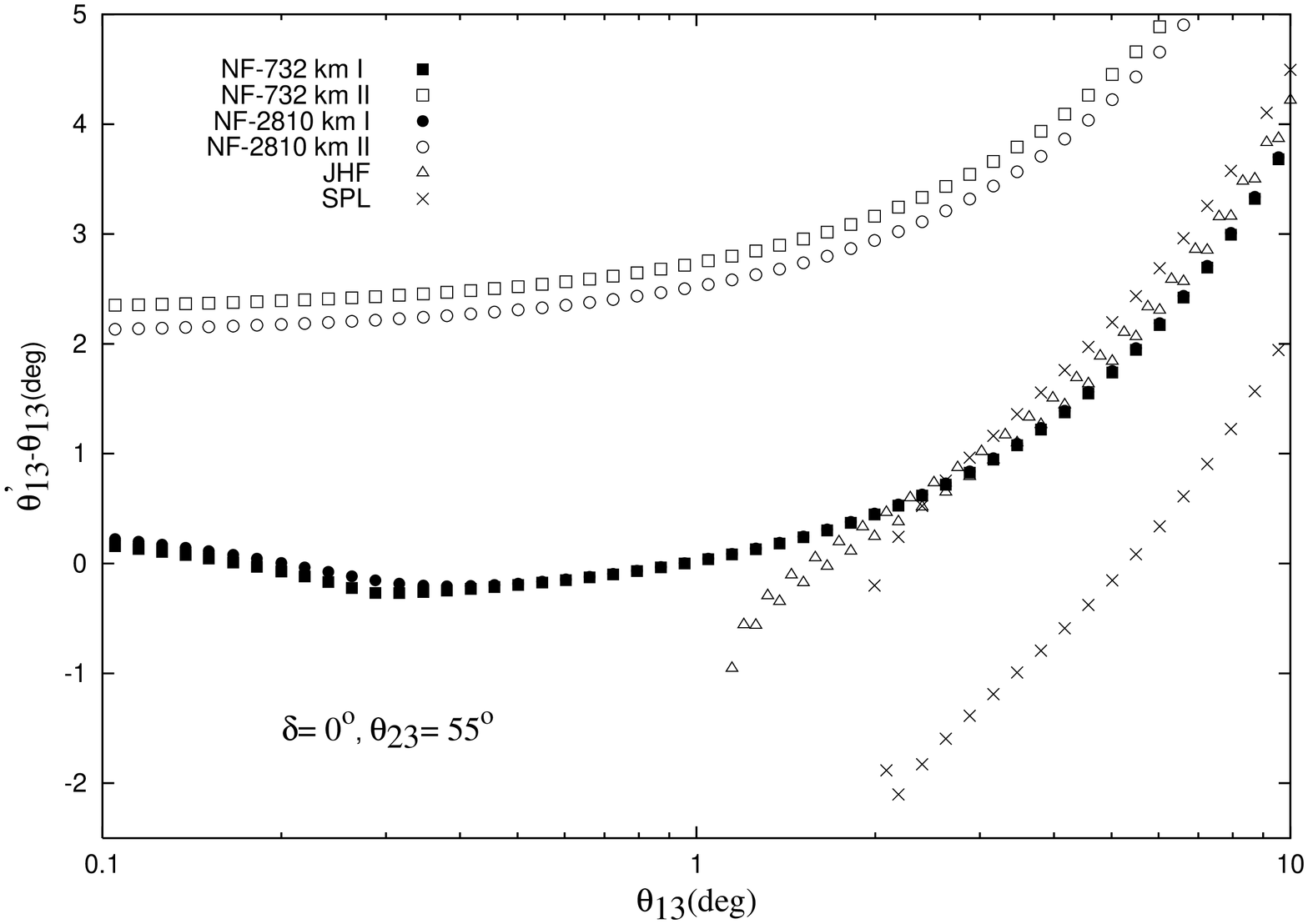, height=8cm,angle=-90} &

\hskip -0.5cm
\epsfig{file=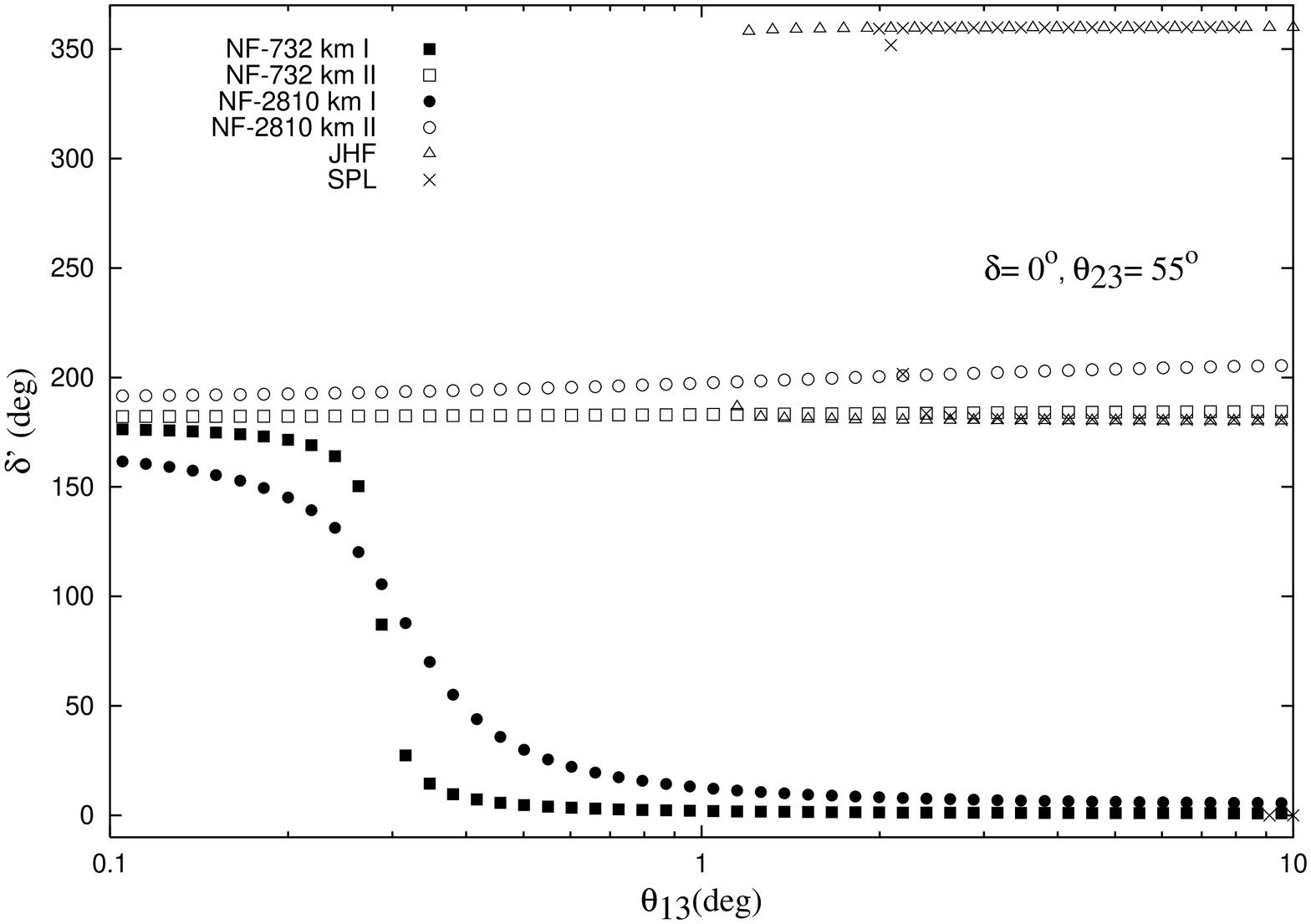, height=8cm,angle=-90} \\
\hskip -0.5cm
\epsfig{file=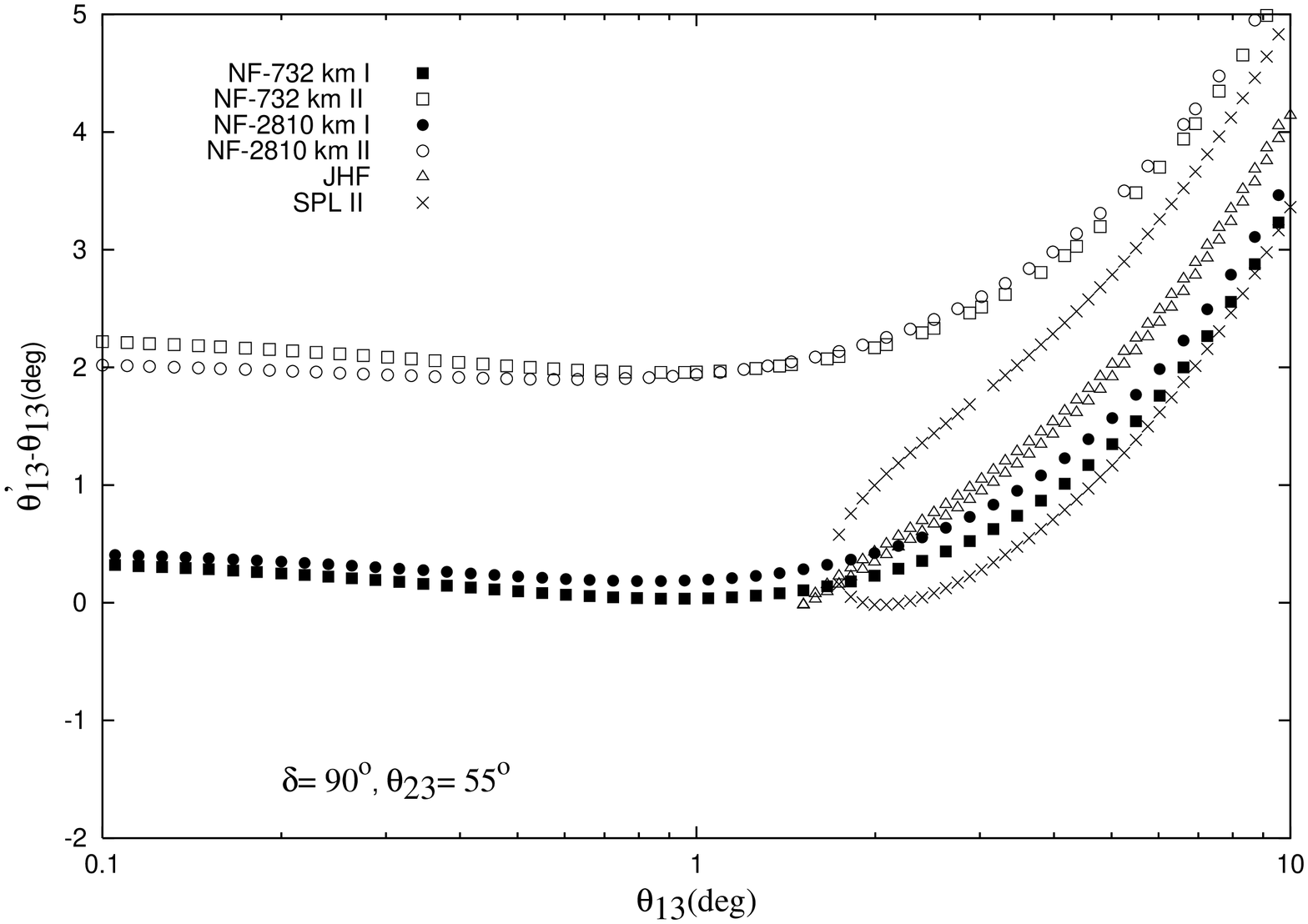, height=8cm,angle=-90} &
\hskip -0.5cm
\epsfig{file=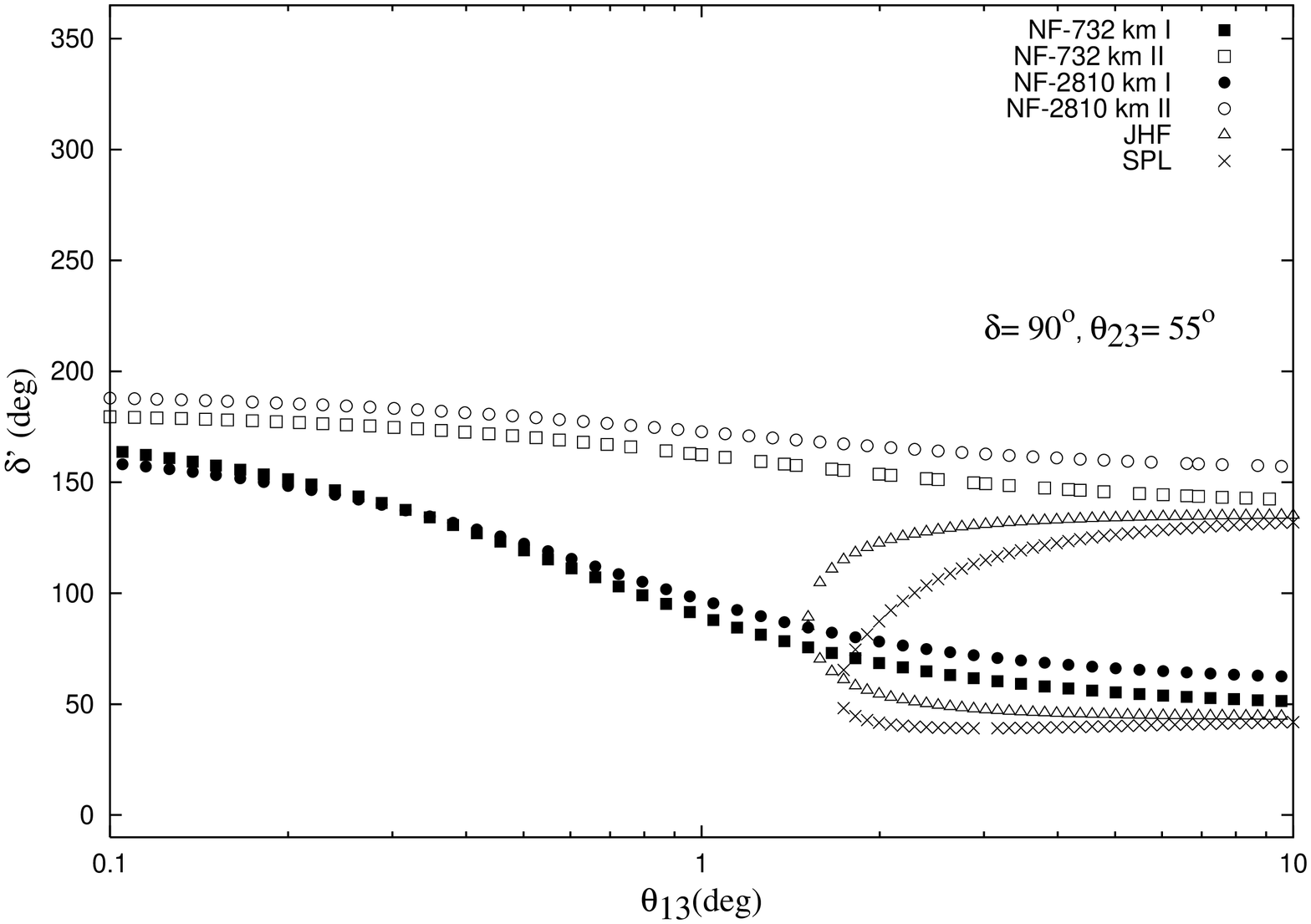, height=8cm,angle=-90} \\
\end{tabular}
\end{center}
\caption{\it{$\tetaotp-\tetaot$ (left) and $\delta'$ (right) for the intrinsic fake solution as a function 
of $\tetaot$, for $\tetatt=55^\circ$, for fixed values of 
$\delta = 0^\circ$(up) and $\delta=90^\circ$ (down).}}
\label{nmm55}
\end{figure}

%%%%%%%%%%%%%%%%%%%%%%%%%%%%%%%%%%%%%%%%%%%%%%%%%%%%%%%%%%%%%%%%%%%%%%%%%%%%%%%
Let us consider in turn the atmospheric and solar regimes. 
For large $\tetaot$, fake $\tetatt$ solutions are given by  
\bea
\sin \delta^{'}&\simeq&\cot \theta_{23}\ \sin \delta, \nonumber\\
\theta^{'}_{13} &\simeq & \tan\theta_{23}\ \theta_{13}+ 
{ \sin 2\theta_{12} \sol \over 2 \sin\left(\atmos\right)}\left(\cos\left(\delta-\atmos\right)-\tan\tetatt \cos\left(\deltap-\atmos\right)\right).
\label{t23atm}
\eea
This system describes two solutions.
For one of them (I) the $L$- and $E$- dependent terms in eqs.~(\ref{t23atm}) 
tend to cancel for $\tetatt \rightarrow \pi/4$, resulting in $\tetaot^{'}= \tetaot$, $\delta^{'}=\delta$ in this limit. 
 The other solution (II) coincides in this limit with that for the 
intrinsic degeneracy, eq.~(\ref{atmburguet}), as 
expected. For both fake $\tetatt$ solutions, deep in the atmospheric regime 
the shift $\tetaotp-\tetaot$ is positive (negative) for $\theta_{23} >(<) \pi/4$.
 Note also that, from eqs.~(\ref{t23atm}), no fake solutions are 
expected for $|\cot \theta_{23} \;\sin \delta| > 1$.
\begin{figure}[t]
\begin{center}
\begin{tabular}{ll}
\hskip -0.5cm
\epsfig{file=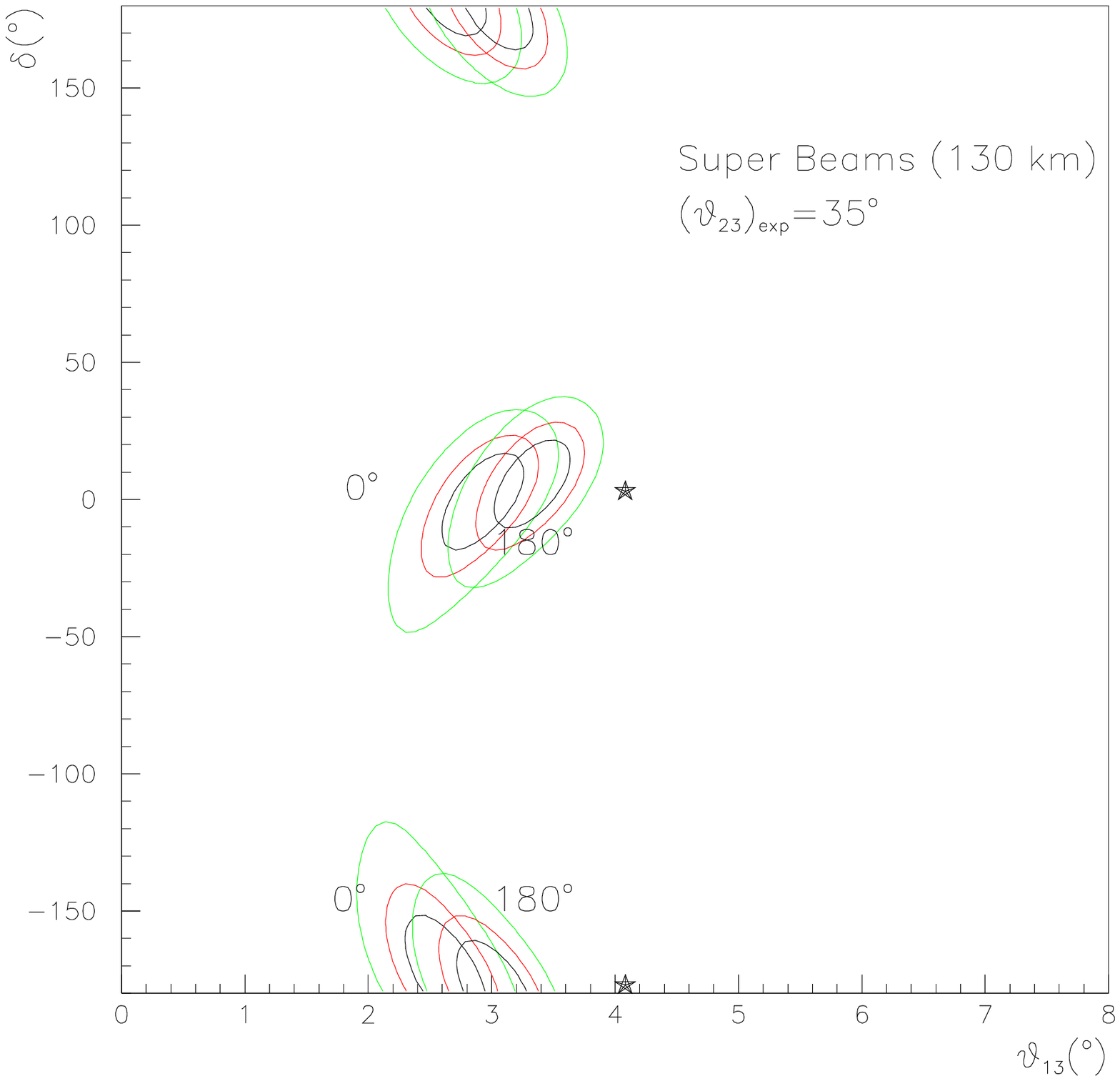, width=8cm} &
\hskip -0.5cm
\epsfig{file=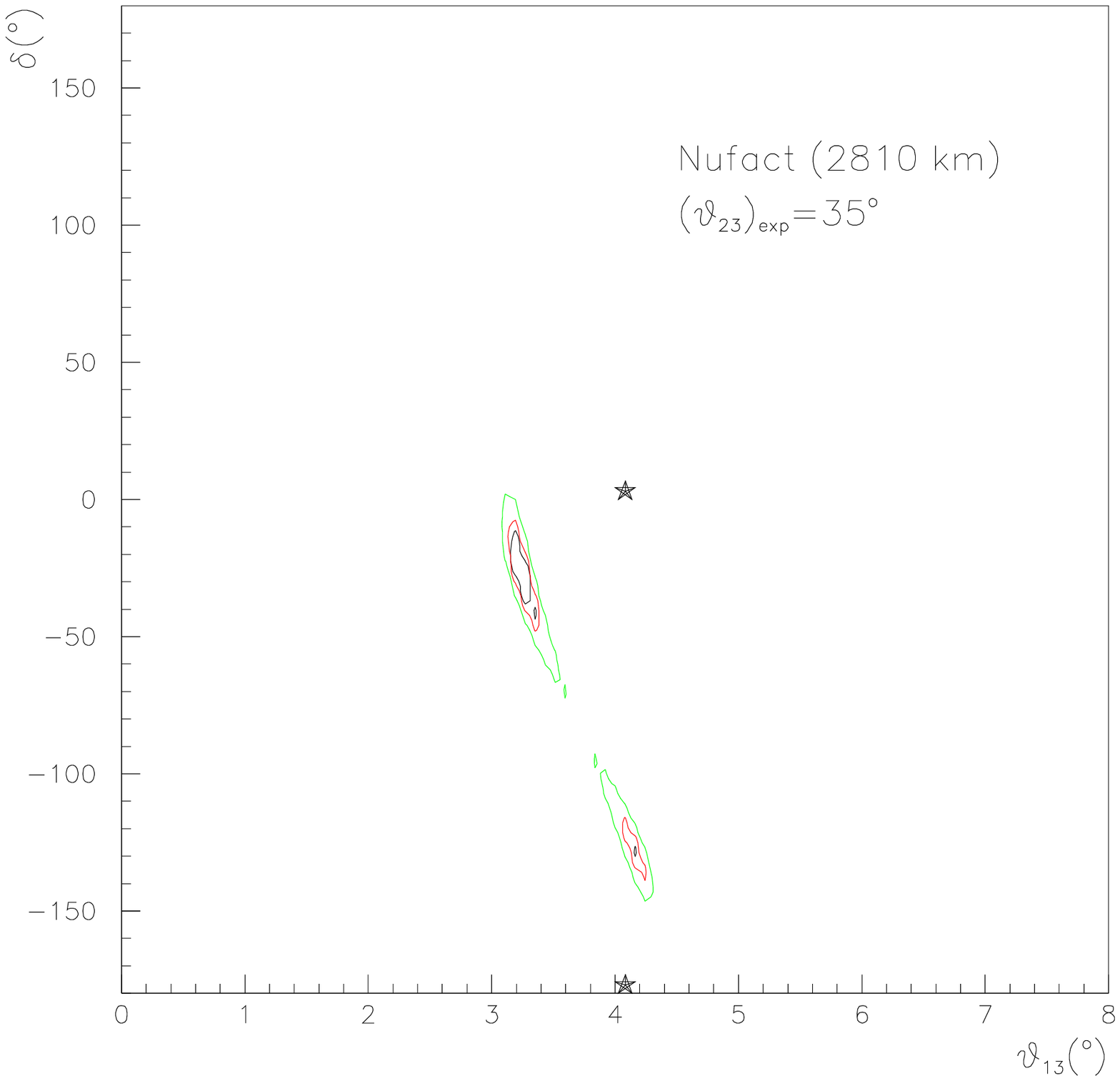, width=8cm} \\
\end{tabular}
\end{center}
\caption{\it{ Fake solutions due to $\tetatt$ degeneracies for SPL-SB results (left) and  a $L=2810$~km NF baseline (right),   
for $\tetatt=35^\circ$, $\theta_{13}=4^\circ$ and  $\delta=- 90^\circ, 0^\circ, 90^\circ, 180^\circ$.
 The combination of the results from both experiments resolves the degeneracies.}}
\label{2810atmosth}
\end{figure}
In the plots of Figs.~\ref{nmm35} and \ref{nmm55}, we show the solutions to eqs.~(\ref{nonequalth}), 
including matter effects, for $\theta_{23}$ at the two extremes of the
$90\%$CL-allowed interval. Note that for large $\tetaot$ there is one solution 
(I) that is more facility-independent than the other, although the 
$E,L$ dependence is sizeable for both solutions (see for instance the curves 
for $\delta=90^\circ$ in Figs.~\ref{nmm55}) when $\theta_{23}$ is so far from maximal.

We have performed fits with the wrong choice of octant for $\tetatt$ and 
central values of $\tetatt$ at the limit of the currently allowed domains. 
The results confirm the expectations above and indicate a situation close to that for the fake sign degeneracies, albeit slightly 
more difficult. 
For instance, at the $L=2810$ km baseline of the NF alone, 
still some fake $\tetatt$ solutions 
remain down to $\tetaot > 2^\circ$, but again they all disappear when combined with the SPL-SB data.  
As an illustration, in Figs.~\ref{2810atmosth} we show the results  for $\tetatt= 35^\circ$ and $\tetaot=4^\circ$,
 at the SPL-SB facility (left) and the $L=2810$ km NF baseline (right).
 The same exercise, but for an $L=732$ km baseline of the NF, results in
 the elimination of the $\tetatt$ degeneracies only for $\tetaot\geq 8^\circ$.

 Let us  now turn to the study of the solar regime.
For $\tetaot\rightarrow 0^\circ$, there are again two fake solutions if the following condition is met:
\bea
\tan^2 \theta_{23} < {1 \over \sin^2\left(\atmos\right)}. 
\eea
Otherwise no solution exists. 
This is important for the larger possible values of $\tetatt$ and well reflected in Figs.~\ref{nmm55},
 which show the exact solutions for $\tetatt = 55^\circ$. Indeed no fake $\tetatt$ degeneracies
 appear in the SB facilities in this case, for $\tetaot$ in the solar regime.

For $\tetaot\rightarrow 0^\circ$, eqs.~(\ref{nonequalth}) can be easily solved to first order 
in $\epsilon_{23} \equiv \tan \theta_{23} - 1$. Solution I becomes in this limit:
\vspace{-0.5cm} \begin{center}
\begin{equation}
\left.\matrix{
\textrm{if}\ \cos 2\theta_{23} \cot\left(\atmos\right)  >0\  \textrm{then} \ \delta^{'}\simeq\ 0 \cr
\textrm{if}\ \cos 2\theta_{23} \cot\left(\atmos\right)    <0\  \textrm{then} \  \delta^{'}\simeq\ \pi}
\right \} \;\;\; \theta^{'}_{13}\simeq \ \sin
2 \theta_{12} \sol \left|\epsilon_{23} \csc\left(\atmosdos\right) \;  \right|.
 \label{t23solarI}
\end{equation}
\end{center}

\begin{figure}[t]
\begin{center}
\begin{tabular}{ll}
\hskip -0.5cm
\epsfig{file=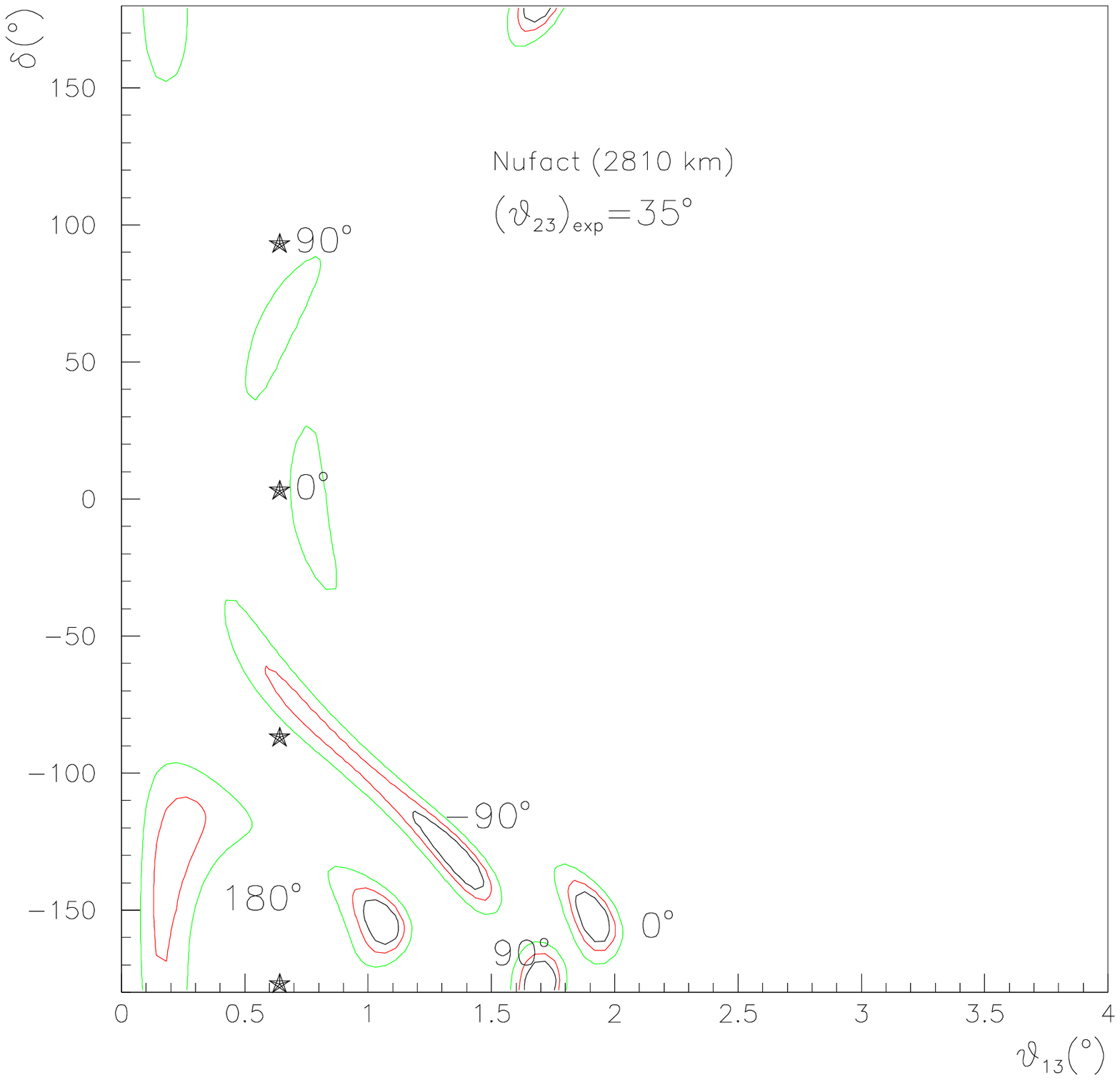, width=8cm} &
\hskip -0.5cm
\epsfig{file=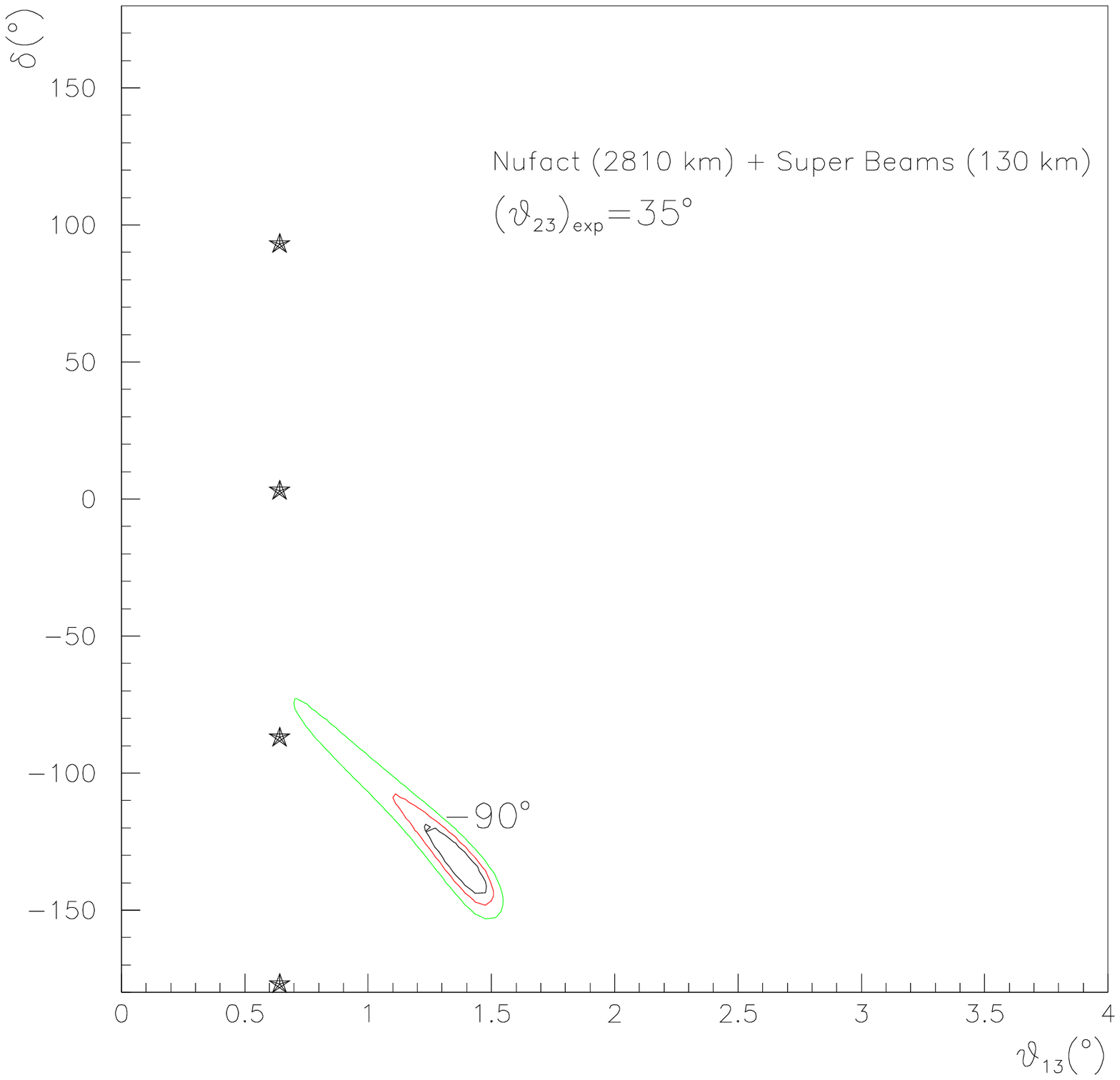, width=8cm} \\
\end{tabular}
\end{center}
\caption{\it{ As Figs.~(\ref{2810solarsgn}) but for the case of $\theta_{23}$ degeneracies.}}
\label{2810solarth}
\end{figure}

Similarly, solution II for $\tetaot \rightarrow 0^\circ$ is given by:
\vspace{-0.5cm}
 \begin{center}
\begin{equation}
\left.\matrix{
\textrm{if}\ \cot\left(\atmos\right)  > 0\  \textrm{then} \ \delta^{'}\simeq\ \pi \cr
\textrm{if}\ \cot\left(\atmos\right)  < 0\  \textrm{then} \  \delta^{'}\simeq\ 0}
\right \} \;\;\; \theta^{'}_{13}\simeq \ \sin 2 \theta_{12}\  \sol  \left( \left|\cot \atmos \right| \pm \epsilon_{23} \cot \atmosdos \right),
 \label{t23solarII}
\end{equation}
\end{center}
where the sign $\pm$ corresponds to the sign($\cot\atmos$). 
The intrinsic degeneracy, eq.~(\ref{solarburguet}), is recovered for $\tetatt=45^\circ$.
Note that, in the solar regime both fake $\tetatt$ solutions have a sizeable  $L,E$ dependence, when 
$\theta_{23}$ is far from maximal.  These two solutions can be seen in 
Figs.~\ref{nmm35} and \ref{nmm55} for small $\tetaot$. Only for the NF setups do solutions I and II remain on the same 
curve in the solar and atmospheric regimes. In the case of the SPL and JHF facilities, they are mixed.

 Figures~\ref{2810solarth} show the fits for $\tetaot=0.6^\circ$, for a NF at $L=2810$ km (left) as well as the same combined 
with the results from the SPL-SB facility (right): only one annoying fake solution remains in 
the latter, which results from the merging of  solution I for SB and solution II for the NF, owing to the finite resolution. See eqs.~(\ref{t23solarI}) and (\ref{t23solarII}).

In general we have found that the NF and SPL-SB combination brings an enormous improvement to the solution of these 
fake degeneracies, particularly 
for large $\tetaot$. The conclusions are rather parallel to those for the fake sign($\Delta m_{13}^2$) solutions, with the 
caveat that for the $\tetatt$ ambiguities,  
solution I, which is harder to resolve, is not that close to satisfying $\sin \delta^{'} =\sin \delta$, and 
it is thus potentially more harmful to the measurement of CP violation. 

As regards the dependence on the solar parameters, the arguments of the previous two sections can be repeated for solutions I and II, when $\theta_{23}$ is close to maximal. When $\theta_{23}$ is farther from 
$\pi/4$, the situation is more confusing since both solutions have a 
dependence on the solar parameters and a detailed exploration of the whole LMA parameter space is necessary.

\section{The silver channels}

One possibility that can help very much to remove degeneracies further is to measure also the 
$\nu_e \rightarrow \nu_\tau$ and  $\bar{\nu}_e \rightarrow \bar{\nu}_\tau$ transition probabilities. 
The relevance of these {\it silver} channels in reducing 
intrinsic degeneracies was studied in ref.~\cite{andrea}, in the atmospheric regime.
Consider the approximate oscillation probabilities\cite{golden,andrea,taus} in vacuum for $\nu_e\rightarrow \nu_\tau$ 
( $\bar \nu_e \rightarrow \bar \nu_ \tau$):
\bea
P_{\nu_ e\nu_\tau ( \bar \nu_e \bar \nu_\tau ) } & = & 
c_{23}^2 \, \sin^2 2 \tetaot \, \sin^2 \left(\atmos\right)  + 
s_{23}^2 \, \sin^2 2 \theta_{12} \, \left( \sol\right)^2 \nn \\
& - & \tilde J \, \cos \left ( \pm \delta - \atmos \right ) \;
 \sol \sin \atmos .
\label{vacexpandetau} 
\eea
They differ from those in eq.~(\ref{vacexpand}) by the interchange $\theta_{23} \rightarrow \pi/2 -\theta_{23}$ and by a change 
in the sign of the interference term. 

 For the intrinsic degeneracies in the atmospheric regime, it follows that the sign 
of $\tetaotp-\tetaot$ will be opposite to that 
for the golden $\nu_e \leftrightarrow\nu_\mu$ ($\bar \nu_e \leftrightarrow \bar \nu_\mu$) channels given in 
eqs.~(\ref{atmburguet}). 
In the solar regime, the intrinsic solutions in these silver channels will thus be identical to eqs.(\ref{solarburguet}) upon 
exchanging $\delta'=0 $ and $\pi$, and the combination of the golden and silver channels remains a promising option.  

 Let us now turn to the fake $\tetatt$ solutions.
When considering 
only $\nu_e\rightarrow \nu_\tau$ and $\bar{\nu}_e\rightarrow \bar{\nu}_\tau$ oscillations,  
the location of the fake solutions related to the $\theta_{23}$ ambiguity, in the atmospheric regime, is:
\bea
\sin \delta^{'}&\simeq&\tan \theta_{23}\ \sin \delta, \nonumber\\
\theta^{'}_{13} &\simeq & \cot \theta_{23}\ \theta_{13} - 
\sin 2\theta_{12}\frac{\sol} {2\sin \atmos}\left(\cos\left(\delta-\atmos\right)-\cot\tetatt \cos\left(\deltap-\atmos\right)\right).
\label{t23atmtaus}
\eea
Thus the shift $\theta'_{13}-\theta_{13}$ at large $\tetaot$ would have the opposite sign to that in eq.~(\ref{t23atm}). 

In the solar regime, on the other hand, solution I for the $\nu_\tau$ appearance 
measurement is the same as that in eq.~(\ref{t23solarI}), while solution II is different, namely:
 \begin{center}
\begin{equation}
\left.\matrix{
\textrm{if}\ \cot\left(\atmos\right)  > 0\  \textrm{then} \ \delta^{'}\simeq\ 0 \cr
\textrm{if}\ \cot\left(\atmos\right)  < 0\  \textrm{then} \  \delta^{'}\simeq\ \pi}
\right \} \;\;\; \theta^{'}_{13}\simeq \ \sin 2 \theta_{12}\  \sol \ \left( \left|\cot \atmos \right| \mp \epsilon_{23} \cot \atmosdos \right).
 \label{t23solarIItaus}
\end{equation}
\end{center}
The condition for the existence of solutions in the solar regime is also different:
\bea
\cot^2 \theta_{23} < {1 \over \sin^2\left(\atmos\right)}. 
\eea
 
A detailed analysis for a realistic experimental 
setup will be done elsewhere \cite{prep}, but we expect that the combination of the two appearance 
measurements: $\nu_e\rightarrow \nu_\mu$ and $\nu_e\rightarrow \nu_\tau$ for both polarities can help to 
resolve the dangerous solution I associated with the $\theta_{23}$ ambiguity, for $\tetaot$ in the atmospheric regime.

Finally, we recall that the disappearance measurements (e.g. $\nu_\mu \rightarrow \nu_\mu$)  should also be helpful in 
reducing these ambiguities for large  $\tetatt$. 
 Obviously, if the angle $\theta_{23}$ will turn out to be close to maximal (as the best-fit point now indicates), the 
$\tetatt$ degeneracies will be of very little relevance.

As for the removal of the fake sign$(\Delta m_{13}^2)$ degeneracies, the {\it silver} channels will also help, for qualitatively 
the same reason as in the combination of facilities with
opposite value of $\cot\atmos$. For maximal $\theta_{23}$,  
the solution of type I  in the silver channel is the same in vacuum  as that
in the golden
channel, and it is thus not expected to disappear in the combination of the 
two appearance measurements.  
The solution of type II, instead, has an opposite displacement in $\theta_{13}$ in the 
atmospheric regime and a difference of $180^\circ$ in the phase in the solar
one. 

\section{Conclusions}

 The extraction of a given set of nature values $(\tetaot, \delta)$ from the detection of neutrino oscillations 
 through the golden channels $\nu_e\leftrightarrow\nu_\mu$ ($\bar{\nu}_e\leftrightarrow\bar{\nu}_\mu$)
 results generically in that the true solution may come out accompanied by  fake ones, which might interfere severely with the 
measurement of CP violation. One of the fake solutions comes from the 
intrinsic correlation between $\delta$ and $\theta_{13}$. The others come 
from the discrete ambiguities: sign($\Delta m^2_{13})$ and sign$\left(\cos 2\theta_{23}\right)$.   

We have shown the enormous potential of combining the data from a superbeam facility and a neutrino factory, to eliminate 
these degeneracies. Because of the sizeable matter effects, neutrino factory baselines that are optimal to measure CP violation (as well as shorter ones), imply 
a considerably smaller ratio $\langle L/E\rangle$ than in the proposed superbeam facilities. 
    It turns out that the location of the fake solutions is very sensitive to this quantity, hence the potential 
of combining the results from both type of facilities.

  We have shown that the fake solutions associated with the sign and $\tetatt$ ambiguities can be grouped in two sets: 
those closer to nature's values (solutions of type I) and those related to the intrinsic fake solution (solutions II).
Generically solutions I show a milder $L/E$ dependence  and are thus more difficult 
to eliminate through this strategy.
We have studied all fake solutions both analytically and through simulations, including realistic background 
errors and efficiencies for a magnetized iron detector at a neutrino factory, and a water Cerenkov one at the proposed 
SPL superbeam  facility.

 The fits have been performed assuming  the LMA-MSW solar solution, with 
$\sin^2 2\theta_{12}\,.\Delta m_{12}^2=10^{-4}$ eV$^2$.
 For $\tetaot$ near its present limit, the combination of the SPL-SB data and those from a short NF 
baseline, i.e. $L=732$ km, is sufficient to resolve all of them and deliver a clean measurement of 
$\tetaot$ and leptonic CP violation.
 With lowering $\tetaot$ but still in the atmospheric regime, although the same setup often produces interesting results,
it is necessary to consider an intermediate NF baseline, i.e. $L=2810$ km, together with the superbeam. In particular the 
sign of $\Delta m^2_{13}$ can be measured from the combination of their data down to $\tetaot > 1^\circ$. 

For values of $0.5^\circ< \tetaot\ < 1^\circ$ most degeneracies still disappear in the combined data from the SPL-SB 
facility and the $L=2810$ km NF baseline, but some fake solutions remain, mainly of type I. While those associated to the sign$(\atmos)$ ambiguity  bias only 
slightly the extraction 
of the true $\tetaot$ and $\delta$ values, those related to $\tetatt$ would remain a problem, if $\tetatt$ were far from maximal.

 A simultaneous error on the assumed sign$\left( \Delta m_{13}^2 \right)$ and $\tetatt$ octant, of course, gives rise to 
additional combined fake solutions: we have checked, for our central values of the oscillation parameters,  that those get resolved 
when the corresponding individual degeneracies get resolved. 
Besides, although we have not done a systematic exploration of the presently allowed
range for the atmospheric and solar parameters, we have argued that we expect
conclusions similar to those obtained in this work, for lower values of $\Delta m^2_{13}$ and $\Delta m^2_{12}$, to the extent that the 
sign of $\cot\left(\atmos\right)$ remains opposite in the two facilities.

  It has previously been pointed out \cite{andrea} that a supplementary  measurement of the {\it silver} 
channels, i.e. $\nu_e\leftrightarrow\nu_\tau$ ($\bar{\nu}_e\leftrightarrow\bar{\nu}_\tau$), could help in removing the intrinsic degeneracy.
We have also discussed in this paper the expected impact of such measurements on resolving the fake sign 
and $\tetatt$ degeneracies. Although a detailed analysis will be done elsewhere,  we expect a big improvement in eliminating in particular 
the dangerous fake solutions associated with the $\tetatt$ octant ambiguity.

 Superbeams and Neutrino Factory are two successive steps in the same path towards the discovery of leptonic CP violation: 
a golden path, not so much for its budgetary cost, but for the solid and shining perspective offered by the combination 
of their physics results.

\section{Acnowledgements}

 We are indebted to Andrea Donini and Stefano Rigolin for useful discussions and comments.
  The work of M.B.~Gavela and J.J.~G\'omez-Cadenas was partially supported by CICYT projects FPA 2000-0980 (GCPA640), and
 SPA2001-1910-C03-02 , respectively. That of O.~Mena and P.~Hern\'andez was partially supported by AEN-99/0692.

\end{document}